\documentclass[aps,twocolumn,superscriptaddress,floatfix,showpacs]{revtex4}
\usepackage{amsmath,euscript,theorem,graphicx,rotating}
\usepackage[letterpaper,hmargin=25mm,vmargin=25mm]{geometry}
\usepackage[latin1]{inputenc}
\usepackage[T1]{fontenc}
\usepackage{txfonts}
\usepackage{url}
\usepackage{array}
\usepackage{xr} 
\usepackage{color}
\usepackage{dcolumn}
\usepackage{bm}
\usepackage{longtable}
\usepackage{xspace}
\usepackage{booktabs}

\usepackage{array}
\newcolumntype{d}[1]{D{.}{.}{#1}}
\newcolumntype{L}{>{$}l<{$}}
\newcolumntype{R}{>{$}r<{$}}
\newcolumntype{C}{>{$}c<{$}}

\newcommand{\rr}{{\mathbf r}}

\newcommand{\RR}{{\mathbf R}}
\newcommand{\rp}{{\mathbf r'}}

\newcommand{\etal}{\emph{et al.}\xspace}

\newcommand{\kJmol}{\unskip\ensuremath{\,\rm{kJ\,mol}^{-1}}\xspace}

\newcommand{\Eexch}[0]{\ensuremath{E^{(1)}_{\rm exch}}\xspace}
\newcommand{\Eindreg}[1]{\ensuremath{E^{(#1)}_{\rm ind,tot}({\rm Reg})}\xspace}
\newcommand{\Eind}[1]{\ensuremath{E^{(#1)}_{\rm ind,tot}}\xspace}

\newcommand{\Eindpol}[1]{\ensuremath{E^{(#1)}_{\rm ind,pol}}\xspace}
\newcommand{\Eindpolreg}[1]{\ensuremath{E^{(#1)}_{\rm ind,pol}({\rm Reg})}\xspace}

\newcommand{\Eindexch}[1]{\ensuremath{E^{(#1)}_{\rm ind,exch}}\xspace}

\newcommand{\deltaHF}[0]{\ensuremath{\delta^{\rm HF}_{\rm int}}\xspace}
\newcommand{\PhiA}[1]{\ensuremath{\Phi^{\mathrm{A}}_{#1}}\xspace}
\newcommand{\PhiB}[1]{\ensuremath{\Phi^{\mathrm{B}}_{#1}}\xspace}
\newcommand{\PhiX}[1]{\ensuremath{\Phi^{\mathrm{X}}_{#1}}\xspace}
\newcommand{\EA}[1]{\ensuremath{E^{\mathrm{A}}_{#1}}\xspace}

\newcommand{\EX}[1]{\ensuremath{E^{\mathrm{X}}_{#1}}\xspace}
\newcommand{\CTreg}[1]{\ensuremath{{\rm CT}^{(#1)}({\rm Reg})}\xspace}
\newcommand{\CTsm}[1]{\ensuremath{{\rm CT}^{(#1)}({\rm SM09})}\xspace}

\newcommand{\CamCASP}{{\sc CamCASP}\xspace}
\newcommand{\ORIENT}{{\sc Orient}\xspace}

\newcommand{\DALTON}{{\sc DALTON} 2.0\xspace}

\newcommand{\Hmat}[2]{\ensuremath{{\sf\bf H}^{(#1)}_{\rm #2}}\xspace}
\newcommand{\wMat}[1]{\ensuremath{{\sf\bf \omega}^{\rm #1}}\xspace}
\newcommand{\amp}[2]{\ensuremath{s^{#1}_{#2}}\xspace}
\newcommand{\ampMat}[1]{\ensuremath{{\sf\bf s}^{#1}}\xspace}
\newcommand{\wB}[2]{\ensuremath{(\omega^{\rm B})^{#1}_{#2}}\xspace}

\mathchardef\lt="313C \mathchardef\gt="313E
\mathcode`<="4268 \mathcode`>="5269
\newcommand{\JCP}[0]{J. Chem. Phys.\ }
\newcommand{\JPCA}[0]{J. Phys. Chem. A\ }

\newcommand{\JPC}[0]{J. Phys. Chem.\ }
\newcommand{\JCTC}[0]{J. Chem. Theory Comput.\ }
\newcommand{\IJQC}[0]{Int. J. Quantum Chem.\ }
\newcommand{\CPL}[0]{Chem. Phys. Lett.\ }
\newcommand{\TCA}[0]{Theor. Chim. Acta\ }

\newcommand{\PRA}[0]{Phys. Rev. A\ }

\newcommand{\PRL}[0]{Phys. Rev. Lett.\ }
\newcommand{\CR}[0]{Chem. Rev.\ }

\newcommand{\MolP}[0]{Mol. Phys.\ }

\newcommand{\JACS}[0]{J. Am. Chem. Soc.\ }

\begin{document}

\title{Charge-transfer from Regularized Symmetry-Adapted Perturbation Theory}

\author{Alston J. Misquitta}
\affiliation{School of Physics and Astronomy, Queen Mary, University of London,
London E1 4NS, UK}

\date{\today}

\begin{abstract}
The charge-transfer (CT) together with the polarization energy appears at
second and higher orders in symmetry-adapted perturbation theory (SAPT).
At present there is no theoretically compelling way of isolating the 
charge-transfer energy that is simultaneously basis-set independent and
applicable for arbitrary intermolecular separation.
We argue that the charge-transfer can be interpreted as a tunneling phenomenon,
and can therefore be defined via regularized SAPT. This leads to a 
physically convincing, basis-independent definition of the charge-transfer energy
that captures subtleties of the process, such as the
asymmetry in the forward and backward charge transfer, 
as well as secondary transfer effects. 
With this definition of the charge-transfer the damping needed for polarization
models can be determined with a level of confidence hitherto not possible.
\end{abstract}



\maketitle

\section{Introduction}
\label{sec:introduction}
Symmetry-adapted perturbation theory (SAPT) remains one of the most
accurate and versatile methods for calculating intermolecular
interaction energies. Within SAPT the interaction energy is decomposed
into physical components such as the electrostatic, exchange-repulsion,
induction and dispersion energies.  This decomposition is useful
both as an interpretative tool and as a means for developing models
for the interaction energy. The latter is made possible as, with
the exception of the short-ranged exponentially decaying
exchange-repulsion energy, each of these components is associated
with a well-defined multipole expansion, the coefficients of which
can be calculated from monomer properties alone.

There is, however, an important exception to the list of physical
components of the interaction energy as described by SAPT: the
charge-transfer energy is not defined as a separate component, but
is included, together with the polarization energy, in the induction
energy at second and higher orders in the interaction operator.
While these components are not usually separated in a SAPT calculation
of the interaction energy, there are approximate methods, such as
that introduced by Stone \cite{Stone93} and used by Stone \& Misquitta
\cite{StoneM09a} in the context of a variant of SAPT based on a
density-functional description of the monomers (termed SAPT(DFT))
\cite{StoneM09a}, which enable us to make a reasonable partitioning
of the induction energy---at least at second-order---into its
charge-transfer and polarization constituents.  Though these methods,
and their supermolecular counterparts \cite{KhaliullinBH-G08},
provide sensible charge-transfer
energies, particularly for complexes at their equilibrium configurations,
they exhibit serious drawbacks, all related to their basis-set
dependence for small intermolecular separations. Furthermore, as
we shall demonstrate in this paper, the Stone \& Misquitta procedure
leads to charge-transfer energies that do not seem to reflect the physical
nature of the process.

There are several reasons why we require a rigorous definition of
the charge-transfer energy.
\begin{itemize}
\item {\em Polarization models}: The polarization energy can be
  unambiguously
  defined through molecular polarizabilities and multipole moments,
  but only for large molecular separations where charge-density
  overlap is negligible.  At shorter separations the expansion must
  be damped, and to determine the damping we require a knowledge
  of the polarization energy, and hence the charge-transfer energy.
\item {\em Validation of density functionals}: One of the remaining
  issues with the exchange functionals used in density functional theory
  (DFT) is their inability to describe the charge transfer
  energy accurately \cite{CohenM-SY08a}. 
  A considerable effort is being made to attempt to remedy
  this problem, but for this to work accurate benchmark charge-transfer
  energies are needed.
\item {\em Energy decomposition analysis (EDA)} techniques are a powerful 
  means of developing an understanding of the nature of molecular
  interactions. However, with the exception of perturbative techniques
  like SAPT, these methods contain an element of ambiguity: energy
  components, particularly the charge-transfer and polarization
  energies, obtained from different EDAs can be significantly
  different.  To benchmark EDA methods we need a physical and
  basis-independent definition of the interaction energy components.
  SAPT already provides such a decomposition for the electrostatic,
  exchange-repulsion and dispersion energies, but not the charge-transfer
  and polarization. This disadvantage needs to be overcome.
\end{itemize}
The second and third points are linked: EDA techniques are used to
determine the physical content of DFT interaction energies, but to
do so reliably, the EDA methods themselves need to be calibrated
against more rigorous decomposition methods, such as SAPT.  In this
paper we will focus on the first and second: the role the charge-transfer
plays in determining accurate polarization models, and the provision
of benchmark charge-transfer energies.

\section{What is the charge-transfer energy?}
\label{sec:what-is-CT}

Intermolecular charge-transfer is a ground-state phenomenon. When
two molecules are in proximity, their electronic charge density can
be delocalised, that is, shared between them, leading to a stabilisation
by an amount that is the charge-transfer energy.  However, from the
point of view of the interacting monomers, the delocalisation can
be viewed in terms of excitations partly localised on the partner 
monomer: these are the
so-called intermolecular charge-transfer excitations. This is why
the charge-transfer energy first appears at second order in
intermolecular perturbation theory.  Common to both views is the
idea of charge delocalisation leading to stability.
One may even suggest that the term `charge-transfer' is misleading,
and rather, `charge-delocalisation' may be a more appropriate term
for this phenomenon.  We will return to this issue of nomenclature
later in the paper.

This simple physical picture of the charge transfer process lies
at the heart of current attempts to define the charge-transfer
energy: if we can calculate interaction energies both by allowing
and suppressing the delocalisation process, the charge-transfer
energy can be defined as an energy difference.

\subsection{Basis-space definitions}
\label{ssec:CT-basis-space-defns}

Currently, definitions of the charge-transfer energy rely on
basis-space localisation methods. There are many flavours of this
kind of approach, all of which are based on the idea: that with
a suitably localised basis set we suppress the charge-transfer-type
excitations, and consequently isolate the charge-transfer from the
polarization energy.  
In the Stone \& Misquitta \cite{StoneM09a}
technique the charge-transfer energy is defined as the difference
in the second-order induction energy, \Eind{2}, calculated in a
dimer centered (DC) and monomer centered (MC) basis set. That is,
\begin{equation}
  \CTsm{2} = \Eind{2}[\rm DC] - \Eind{2}[\rm MC].
  \label{eq:CT-SM09}
\end{equation}
Here, unless otherwise specified, the induction energy will be 
assumed to be the sum of the polarization and exchange components
as \cite{MisquittaS08a}: $\Eind{n} = \Eindpol{n} + \Eindexch{n}$.
In the MC basis the molecule is described using basis functions
located on its own nuclei only, therefore suppressing any excitation
that could give rise to an increased electronic density on the
partner monomer. That is, charge-transfer would be suppressed.  By
contrast, in the DC type of basis, the monomers are additionally
described using basis functions located on their partners, therefore
allowing charge-transfer type excitations. Hence we are led to
the above definition of the CT energy which we will henceforth
term CT-SM09.

Note that in practical calculations, instead of the DC type of basis
we often use an equivalent, but smaller, MC+ type of basis which includes
mid-bond and the so-called far-bond functions \cite{WilliamsMSJ95}. 
The far-bond functions are a subset of the basis of the partner
monomer; usually the s- and p-functions only. Extensive tests
have shown that these two types of basis set result in nearly
identical induction energies \cite{WilliamsMSJ95}.
Consequently we will use the 
DC and MC+ types interchangeably in eq.~\eqref{eq:CT-SM09} and will
consider induction energies calculated in either basis type
to include all CT effects.

There are, however, a few objections that one could raise with the
CT-SM09 definition:
\begin{itemize}
\item Equation~\eqref{eq:CT-SM09} leads to a basis-dependent definition of the
  CT energy. As the monomer basis gets larger and more complete, the distinction between
  the MC and DC basis types grows smaller. This would lead to smaller apparent 
  CT energies. This issue has been acknowledged by Stone \& Misquitta who 
  have argued that this effect is indeed present, but is small enough that for
  practical basis sets this is not and issue.
\item A more subtle, but related shortcoming of eq.~\eqref{eq:CT-SM09} is that
  it inevitably leads to a separation-dependent definition of the charge-transfer.
  As the intermolecular separation reduces, the MC-type basis set is better able to 
  describe CT-type excitations, that is, the definition of the CT gets
  progressively worse.
\item The CT-SM09 definition is really only the second-order charge-transfer
  energy. There are contributions to the CT from third  and higher orders
  in perturbation theory that, as we shall see, are comparable to the
  second-order CT. 
\item This definition relies on the use of localised atomic basis sets
  and offers no clear route to extension to plane-wave basis sets.
  Consequently, eq.~\eqref{eq:CT-SM09} is largely limited to programs that use 
  Gaussian-type orbitals. This is certainly not a significant issue, but
  it would be advantageous to develop a method that was independent of the
  type of basis set.
\end{itemize}

\begin{figure}
  \includegraphics[width=0.5\textwidth,clip]{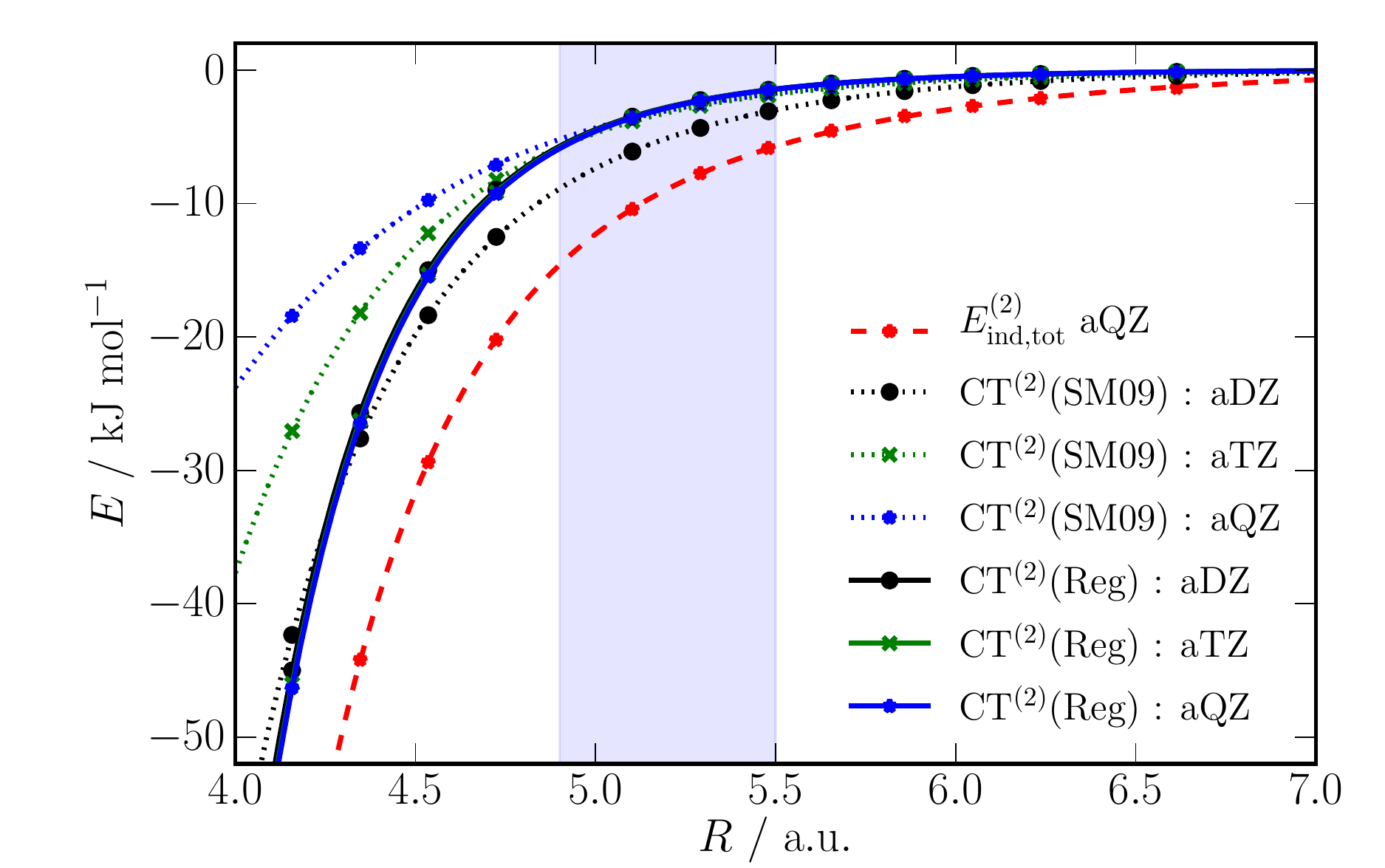}
  \caption[H2O2: Variation of CT-SM09 with basis set]{
  The second-order charge-transfer for the water dimer calculated using the 
  Stone--Misquitta procedure (CT-SM09) with three basis sets (dotted lines).
  Also included is that total second-order induction energy \Eind{2} calculated
  using the aug-cc-pVQZ basis of the MC+ type.  \Eind{2} calculated with the 
  aug-cc-pVDZ and aug-cc-pVTZ MC+ basis types are visually identical and
  are not shown. $R$ is the O$\cdots$O separation. The shaded region
  indicates the range of $R_{\rm OO}$ values attained in the water dimer
  (largest value) and clusters of water through to the water hexamer 
  (smallest value). 
  \label{fig:water2-CT-SM09-var-basis}
  }
\end{figure}

The first two points are illustrated in
fig.~\ref{fig:water2-CT-SM09-var-basis} for the water dimer in its
minimum-energy, H-bonded conformation. The variation of the CT-SM09
energy with basis set is indeed small at the equilibrium separation,
particularly for the two larger basis sets.
However, as the intermolecular separation reduces,
the differences between all basis sets grows, with the aug-cc-pVQZ
basis resulting in the least amount of (second-order) charge-transfer
energy at short separations. 

These short separations are important for several reasons.  The
separation of water molecules can decrease significantly in clusters
due to a combination of many-body polarization and charge-transfer
effects. For example, the oxygen--oxygen separation in the water
dimer at its global energy minimum configuration is $2.9$ \AA, but
it is less than $2.5$ \AA\ in the cage hexamer. This range is indicated
in fig.~\ref{fig:water2-CT-SM09-var-basis}. Presumably the intermolecular
separation could be even smaller for larger clusters with stronger
many-body effects, or for water under pressure.

The differences in CT-SM09 calculated using the two larger basis
sets may seem small even at the short end of the $R_{\rm OO}$ range,
but the variation in CT-SM09 with basis at even shorter
separations is large enough that the polarization 
damping model---which is fitted largely to these energies---becomes basis-set
dependent with damping parameters varying from $1.4$, to $1.7$, through
$1.9$ a.u.\ for the aDZ, aTZ and aQZ basis sets. Not only is this
variation large, but the damping parameters tend to depend strongly on
the range of data used---a possible indication that the energies
we are fitting to are not entirely polarization, and might be
contaminated with charge-transfer. This variation in the damping
parameter has no effect on the two-body interaction energy, but
causes the many-body polarization energy for clusters of water
molecules to vary considerably.  For example, with the above damping
parameters, the many-body polarization energy of the cyclic water
pentamer varies from $-59$, to $-73$, to $-79$ \kJmol. This variation is
large enough to make these polarization models unreliable and almost
unusable without explicitly fitting them to reproduce the many-body
energies of such clusters \cite{SebetciB10a}. While this can be done,
this approach is tedious and unsatisfactory as there is always the 
possibility that a damping models that works for one set of clusters
may not work for another.

The Stone and Misquitta technique is but one of many methods that
attempt to define the charge-transfer energy. 
Before continuing we mention two of these:
Schenter and Glendening \cite{SchenterG96a} have proposed a 
decomposition based on natural bond orbitals, which has the merit
that the charge-transfer energies appear to converge with basis set,
but these energies  are unreasonably large.
More recently Khaliullin \etal \cite{KhaliullinBH-G08} developed
a method that operates on a principle similar to that of Stone and Misquitta, 
only this time absolutely localised molecular orbitals (ALMOs) to
suppress charge-transfer-type excitations.
In a recent analysis by Azar \etal \cite{AzarHSH-G13} 
the ALMO method has been demonstrated to exhibit shortcomings 
similar to those of the Stone and Misquitta definition: there is no
complete basis set limit to the charge-transfer energy defined in 
this procedure.

\section{Charge-Transfer via regularized SAPT}
\label{sec:CT-regSAPT}

The Stone--Misquitta and related approaches are all basis-space methods;
that is, these methods attempt to isolate the charge-transfer energy 
by manipulating the atomic or molecular basis sets to try to suppress
CT-type excitations. To see how we may find an alternative approach 
consider the Rayleigh-Schr\"{o}dinger perturbation expression for the
second-order induction energy of molecule A (exchange effects
are included in a separate term):
\begin{align}
  \Eindpol{2}(A) & =  {\sum_{r \ne 0}}
      \frac{|< \PhiA{0}\PhiB{0} | \hat{V} | \PhiA{r}\PhiB{0} > |^2}
           { \EA{0} - \EA{r} }  \nonumber \\
                 & = {\sum_{r \ne 0}}
      \frac{|< \PhiA{0} | V^{\rm B} | \PhiA{r} > |^2}
           { \EA{0} - \EA{r} }  \nonumber \\
                 & = < \PhiA{0} | V^{\rm B} | \PhiA{0}(1) > 
                   \label{eq:E2indpol},
\end{align}
where the \PhiX{n} and \EX{n} are the exact eigenstates and eigenvalues of
monomer X=A,B and $V_{B}$ is the total electrostatic potential of
monomer B, that is, it arises from both electrons and nuclei and can be
written as
\begin{align}
  V^{\rm B}(\rr) = - \sum_{\beta} \frac{Z_{\beta}}{|\rr-\RR_{\beta}|}
              + \int \frac{\rho^{\rm B}(\rp)}{|\rr-\rp|} d\rp,
          \label{eq:VB}
\end{align}
where $Z_{\beta}$ and $\RR_{\beta}$ are the nuclear charges and 
position vectors of monomer B and $\rho^{\rm B}$ is the un-perturbed
electronic charge-density of B.
In the last step in eq.~\eqref{eq:E2indpol} we have defined the
first-order induction wavefunction correction $\PhiA{0}(1)$:
\begin{align}
  \PhiA{0}(1) = {\sum_{r \ne 0}}
      \frac{\PhiA{r} < \PhiA{r} | V^{\rm B} | \PhiA{0} >}
           { \EA{0} - \EA{r} }.
       \label{eq:firstord-wavefn-ind}
\end{align}
Similar expressions exist for the second-order induction energy of 
monomer B.
The sub-script `pol' in eq.~\eqref{eq:E2indpol} indicates that this is the
energy obtained from the so-called polarization expansion \cite{JeziorskiMS94},
that is, without the inclusion of exchange effects.
This notation is unfortunate as it can lead to the erroneous conclusion
that \Eindpol{2}(A) is the polarization energy. 

\begin{figure}
  \includegraphics[width=0.49\textwidth,bb=0 0 900 300,clip]{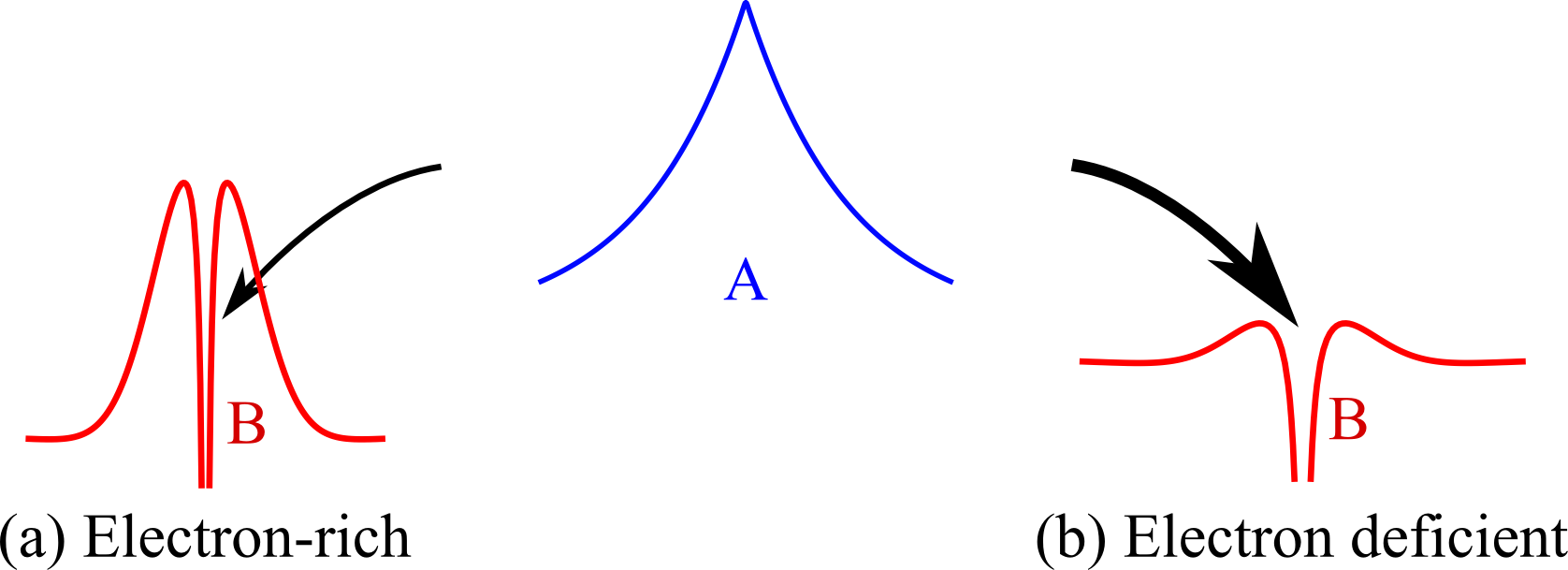}
  \caption[Illustration of CT process]{
  Interpreting the charge-transfer as a tunneling process. We can identify
  two cases for atom B: (a) Electron-rich: The nuclear potential of B is 
  heavily screened by its electronic density, and (b) Electron-deficient: The
  nuclear potential of B is poorly screened.
  Charge-transfer from the electronic density of A (centre, blue)
  into the potential of B is minimal in case (a), but it is significant
  for the case (b). This is illustrated by the width of the arrows.
  \label{fig:CT-tunneling}
  }
\end{figure}

The third form of \Eindpol{2}(A) in eq.~\eqref{eq:E2indpol} allows us to make
the useful interpretation of the second-order induction as the second-order
energy response to the static potential of the partner monomer (or, more
generally, the potential arising from the entire environment of 
monomer A). This potential, eq.~\eqref{eq:VB}, consists of two
parts: an attractive, but singular, term arising from the nuclei of B,
and a finite, repulsive term arising from the electrons of B.
At long-range, monomer A sees the nett effect of these two terms, but 
at short-range, when charge-densities overlap, monomer A 
additionally responds to the singularity arising from the nuclear potential
of B. The electronic density of B screens the singularity of the 
nuclear potential, but the extent of the screening depends on whether
B is electron-rich or electron-deficient. For an electron-rich atom B
(such as the electronegative oxygen atom in water) the nuclear potential
is heavily screened, leading to little tunneling of the electronic
charge density of A into the nuclear potential of B. However, for an 
electron-deficient atom (such as the electropositive hydrogen atom in
a water molecule) the screening is weak, leading to a significant 
degree of tunneling. 

This interpretation of the charge-transfer offers us to a natural way of
suppressing the charge-transfer process altogether: if the potential well
developed by the nuclear potential can be suitably eliminated, there
would be no possibility of tunneling, and consequently no charge 
transfer. This can be achieved by {\em regularizing} the nuclear 
potential that appears in eq.~\eqref{eq:VB} to suppress the nuclear well.

\subsection{Regularization}
\label{ssec:regularization}
Regularization is a technique originally developed to accelerate the 
convergence of the class of SAPT theories. Here we will be concerned
mainly with the version of SAPT based on 
symmetrized Rayleigh-Schr\"{o}dinger (SRS) perturbation theory.
The subject of regularization and the role it plays in convergence
of SAPT is comprehensively discussed in Refs. \onlinecite{PatkowskiKJ01}
and \onlinecite{PatkowskiJS04a}. Although a full discussion of convergence
issues would be out of place in this article, we will discuss some of the
important issues here as these will serve to place the concept of 
regularization in context.

Possibly the earliest major attempt to understand the convergence properties of
intermolecular perturbation theories was made by Claverie \cite{Claverie71a}
who argued that the polarization expansion \cite{JeziorskiMS94} 
(essentially many-body Rayleigh-Schr\"{o}dinger perturbation theory)
on which SRS theory is based either diverged, or if it did converge, 
for systems with more than two electrons, it would converge to a 
Pauli-forbidden state which would be more strongly bound than the 
physical ground state. These ideas were expanded by Kutzelnigg
\cite{Kutzelnigg80a} who showed that the polarization expansion indeed
diverged. Simultaneously, Morgan and Simon \cite{MorganS80} proved that
if any of the interacting atoms had atomic number greater than two, the 
physical ground state of the system could be buried in a continuum of 
Pauli-forbidden states.

The cause of the divergence of the polarization
expansion, and consequently SRS theory \cite{Adams99a}, lies in the 
presence of the unphysical states. Adams explored this issue in a series
of papers (see for example Ref.~\onlinecite{Adams99a}) and showed
that the concept of regularization introduced by Herring \cite{Herring62a}
could be used to de-stabilize these un-physical tunneling states
and hence to ensure the convergence of the theory. 
Simultaneously, Patkowski and collaborators \cite{PatkowskiJS01a,PatkowskiJS04a}
applied similar ideas to develop regularized perturbation theories that 
converged to the physical ground state of the dimer. The idea here is
to write the singular electron--nuclear potential as a short-ranged,
singular part and a long-ranged part that is well-behaved. 
In the notation used by Patkowski \etal this is expressed as
\begin{align}
  \frac{1}{r} = v_p(r) + v_t(r),
  \label{eq:Vreg}
\end{align}
where $v_t$ is the singular, short-ranged part and $v_p$ the long-ranged,
well-behaved part of the nuclear potential. Various schemes can be used
to achieve this splitting, here we will use the Gaussian-based
scheme \cite{PatkowskiJS01a}:
\begin{align}
  v_p(r) & = \frac{1}{r}\left( 1 - e^{-\eta r^2} \right), \nonumber \\
  v_t(r) & = \frac{1}{r} e^{-\eta r^2}.
    \label{eq:gaussian-reg}
\end{align}
Patkowski \etal used SRS theory for the non-singular 
part ($v_p$) and a strong symmetry-forcing theory (using a symmetrization
technique completely suppressing the Pauli-forbidden terms) for the 
singular part \cite{PatkowskiJS01a,Adams02a}, leading to a unified theory
proven to be convergent for the van der Waals and as well as chemical bonding
separations\cite{PatkowskiJS04a}.

\subsection{Charge-transfer via regularization}
\label{ssec:CT-reg}
The issues discussed above are fundamental to foundation and development
of perturbation theories, but are not germane at low orders in perturbation
theory where even weak symmetry-forcing theories like SRS theory
are known to yield sensible results \cite{JeziorskiMS94,PatkowskiJS04a,Adams02a}.
There is a tremendous body of work demonstrating the accuracy of low-order
SAPT (truncated at second or third order with higher-order corrections 
estimated in a non-perturbative manner) for numerous systems
(for example see Refs.~\onlinecite{JeziorskiS98,JeziorskiS02,MisquittaPJS05b,
PodeszwaBS06b,BukowskiSGvanderA06}). For such a truncated theory regularization
has a very different role to play: following on from our discussion of 
the charge-transfer, we argue that, if used at second and higher orders
in perturbation theory, the regularization procedure can be used to de-stabilize
the charge-transfer states and consequently allow us to define a charge-transfer
free interaction energy.

At second order in perturbation theory we calculate the regularized 
induction energy, $\Eindreg{2}$, that, with an appropriate
amount of regularization that is yet to be defined, is free of charge-transfer 
contributions (details below). 
Then, in a manner analogous with eq.~\eqref{eq:CT-SM09}, we may define
the second-order charge-transfer energy as
\begin{align}
  \CTreg{2} = \Eind{2} - \Eindreg{2}.
  \label{eq:CT-Reg}
\end{align}
Here $\Eindreg{2}$ is the regularized second-order induction energy that
may be identified with the true polarization energy. 
There is no basis restriction on the above definition, except that the basis 
set used needs to be large enough to converge the total induction energy, 
which is the usual requirement for any energy calculation. 

\subsection{Implementation}
\label{ssec:implementation}

We have implemented a version of regularized SRS theory (R-SRS) derived
by Patkowski \etal in Ref.~\cite{PatkowskiSJ-RegInd}.
In R-SRS theory the dimer wavefunction corrections are obtained in response
to the interaction operator with regularized nuclear potentials, but the
interaction energy corrections are calculated using the original,
un-regularized interaction operator. 
For the second-order induction energy this means that instead of 
using $\PhiA{0}(1)$, we calculate first-order induction wavefunction
correction in response to the {\em regularized} electrostatic
potential $V^{\rm B}_{\rm Reg}$:
\begin{align}
  V^{\rm B}_{\rm Reg}(\rr) = - \sum_{\beta} Z_{\beta}v_p(\rr-\RR_{\beta})
              + \int \frac{\rho^{\rm B}(\rp)}{|\rr-\rp|} d\rp,
          \label{eq:VB-reg}
\end{align}
to give
\begin{align}
  \PhiA{0}(1)[{\rm Reg}] = {\sum_{r \ne 0}}
      \frac{\PhiA{r} < \PhiA{r} | V^{\rm B}_{\rm Reg} | \PhiA{0} >}
           { \EA{0} - \EA{r} }.
       \label{eq:firstord-wavefn-ind-reg}
\end{align}
The regularised second-order polarization component of the induction energy
is then defined as:
\begin{align}
  \Eindpolreg{2} = < \PhiA{0} | V^{\rm B} | \PhiA{0}(1)[{\rm Reg}] >.
\end{align}
To this, as always, we need to add the similarly regularized exchange-induction
energy \cite{JeziorskiMS94,PatkowskiJS04a,Adams02a}.
Notice that in this step we have used the full electrostatic potential
of monomer B.

The SAPT(DFT) expression for the polarization part of the induction
energy of monomer A in response to the field of B is
\begin{align}
  \Eind{2}({\rm A \leftarrow B}) = 2 \amp{i}{v} \wB{v}{i}
\end{align}
where $i$ and $v$ label occupied and virtual states, 
$\wB{v}{i}$ are matrix elements of the unperturbed potential of 
monomer B given in eq.~\eqref{eq:VB}, and the amplitudes $\amp{i}{v}$ 
are obtained, in the case of SAPT(DFT), by solving the coupled Kohn--Sham
equations
\begin{align}
  \Hmat{1}{A} \ampMat{A} = - \wMat{B}
    \label{eq:amplitudes-LRDFT}
\end{align}
where $\Hmat{1}{A}$ is the electric Hessian \cite{Casida95,ColwellHL96} of
Kohn--Sham linear-response theory that is given in full form for 
hybrid functionals in Ref.~\onlinecite{MisquittaS08a}.
The expression for $\Eindexch{2}({\rm A \leftarrow B})$ also involves the 
amplitudes defined above, though the matrix elements multiplying it
are more complex and are given in full form in 
Refs.~\onlinecite{SAPT-METECC,PatkowskiSJ-RegInd}.
Analogous expressions exist for the induction energy of monomer B due to the
field of A.

In regularized SAPT(DFT) (R-SAPT(DFT)) the amplitudes --- the coefficients
of the first-order induction wavefunction --- are calculated using 
eq.~\eqref{eq:amplitudes-LRDFT}, but this time with the matrix elements on the
R.H.S.\ replaced by those calculated with the regularized electrostatic 
potential $V^{\rm B}_{\rm Reg}(\rr)$ given in eq.~\eqref{eq:VB-reg}.
Otherwise, the expressions for $\Eindpol{2}$ and 
$\Eindexch{2}$ are identical with the non-regularized forms.
These equations have been implemented in the \CamCASP program \cite{CamCASP}
and are available as part of a regular SAPT(DFT) calculation of the 
interaction energy.
Note that the expression for $\Eindexch{2}$ used in this paper does not
involve scaling \cite{MisquittaS08a}. The scaling expression seems to work
by cancellation of errors (of the second-order terms with corresponding 
terms in the \deltaHF estimate of the higher-order induction effects 
\cite{JeziorskiMS94}) 
and cannot be relied upon to give reasonable results for the regularized
induction energy.

\section{Numerical details}
\label{ssec:numerical}

All calculations of SAPT(DFT) energies and molecular electrostatic and polarization
models have been performed with the \CamCASP \cite{CamCASP} program. 
The Kohn--Sham orbital orbitals and orbital energies used in by the \CamCASP 
program were obtained using the \DALTON \cite{DALTON2} program with a patch 
distributed as part of the SAPT2008 \cite{SAPT2008} suite of codes. 

SAPT(DFT) induction energies were calculated using the PBE0 \cite{PerdewBE96,AdamoB99a}
exchange-correlation functional asymptotically corrected (AC) using the Fermi--Amaldi
\cite{FermiA34} long-range form and the Tozer--Handy splicing function
\cite{TozerH98}. For details see Ref.~\onlinecite{MisquittaPJS05b}.
The second-order induction energies were evaluated using the
hybrid form of the linear-response kernel \cite{MisquittaPJS05b,MisquittaS08a}.
\Eindexch{2} was evaluated without scaling \cite{MisquittaS08a} using the 
expressions in Refs.~\onlinecite{HesselmannJS05,PodeszwaBS06a}. This was found
to be necessary for the regularized induction energies.

Distributed multipoles have been calculated using the GDMA2 \cite{Stone05} module
that is part of the \CamCASP suite. Unless otherwise stated, these include
terms to rank 4 (hexadecapole moments) on all sites, including the hydrogen atoms.
Distributed (anisotropic) polarizabilities have been calculated using the 
Williams--Stone--Misquitta (WSM) method \cite{MisquittaS08a,MisquittaSP08} 
with terms to rank 3 on all sites except for the hydrogen atoms for which 
these were restricted to rank 1. 
Molecular multipole and polarizability models have been calculated using
a d-aug-cc-pVTZ basis with the PBE0/AC density and density response functions.
Model energy evaluations were performed using the \ORIENT \cite{Orient4.6} 
program.

\section{Results}
\label{sec:results}

\subsection{Determining the regularization parameter $\eta$}
\label{ssec:eta}
Regularization involves the parameter $\eta$ that has the units of 
$L^{-2}$. Equivalently, ${\eta}^{-1/2}$ has the dimensions of length.
From fig.~\ref{fig:CT-tunneling} and the discussion in Sec.\ \ref{sec:CT-regSAPT}
we know that this length-scale will be associated with the width of the 
screened nuclear potential, which, in principle, will be dependent on the
atom type and bonding environment. We are faced with two choices: either the
parameter $\eta$ is only weakly dependent on atom type, in which case a 
fixed value may be used for all calculations, or this parameter exhibits a
strong atom-type dependence, in which case the regularization procedure will
be potentially cumbersome to apply in practice. The first and most important
question is how are we to determine the appropriate value of $\eta$? 

In principle, $\eta$ could be obtained by examining the screened nuclear 
potential and determining the appropriate regularization needed to 
`fill' it in to as to prevent all tunneling states. But it is at present
unclear what would constitute sufficient `filling', consequently
we have instead opted for two alternate procedures:
\begin{itemize}
  \item We could determine $\eta$ by observing that at
  the optimal regularization there will be no charge-transfer but the polarization
  component of the induction energy will be left unchanged. Consequently, 
  all basis sets large enough to describe the pure polarization effect would
  result in the same regularized induction energy at all separations.
  \item Alternatively, we could determine $\eta$ by requiring that all
  the induction energy in rare gas dimers is charge-transfer.
\end{itemize}

\begin{figure}

  \includegraphics[width=0.5\textwidth]{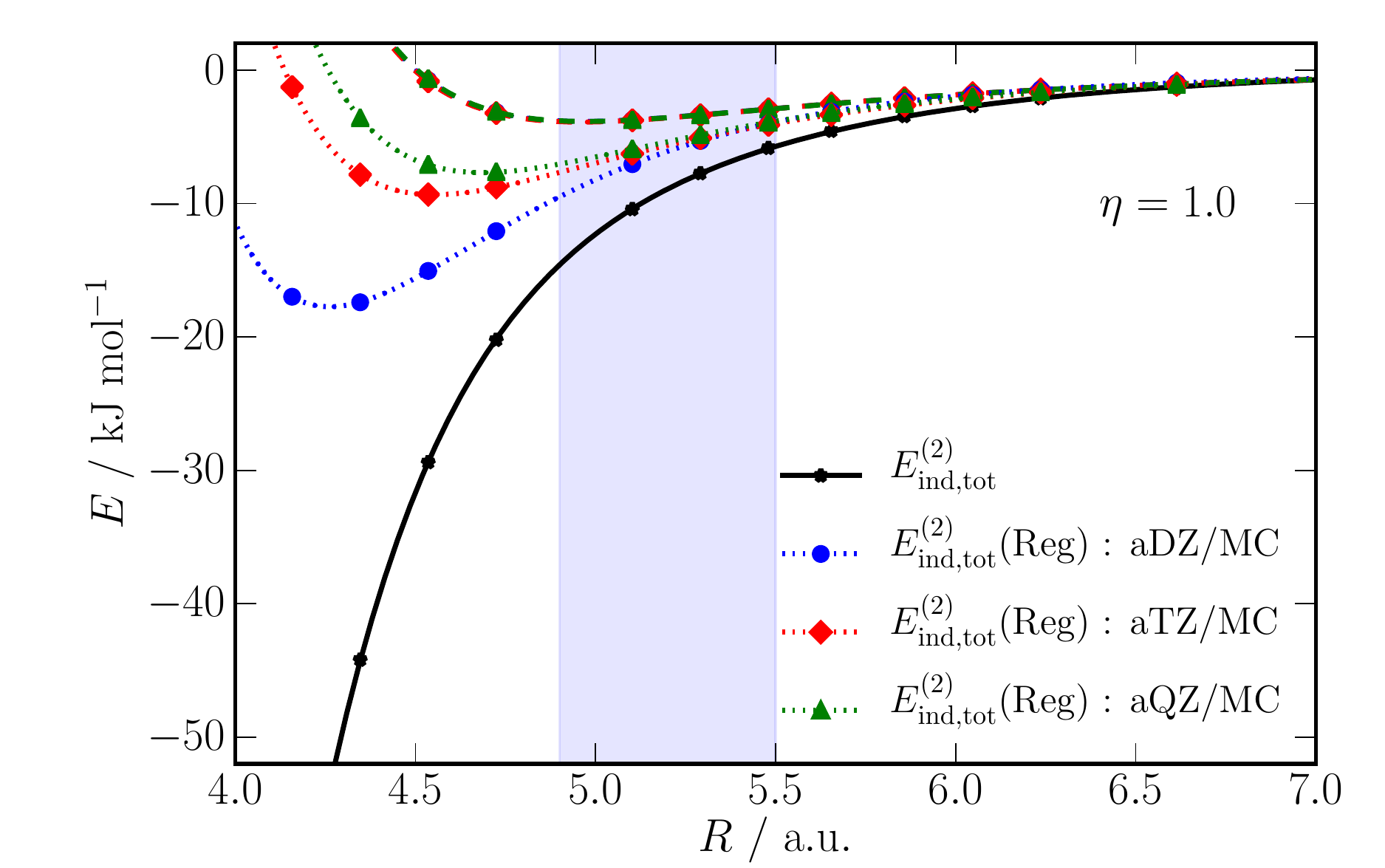}
  \includegraphics[width=0.5\textwidth]{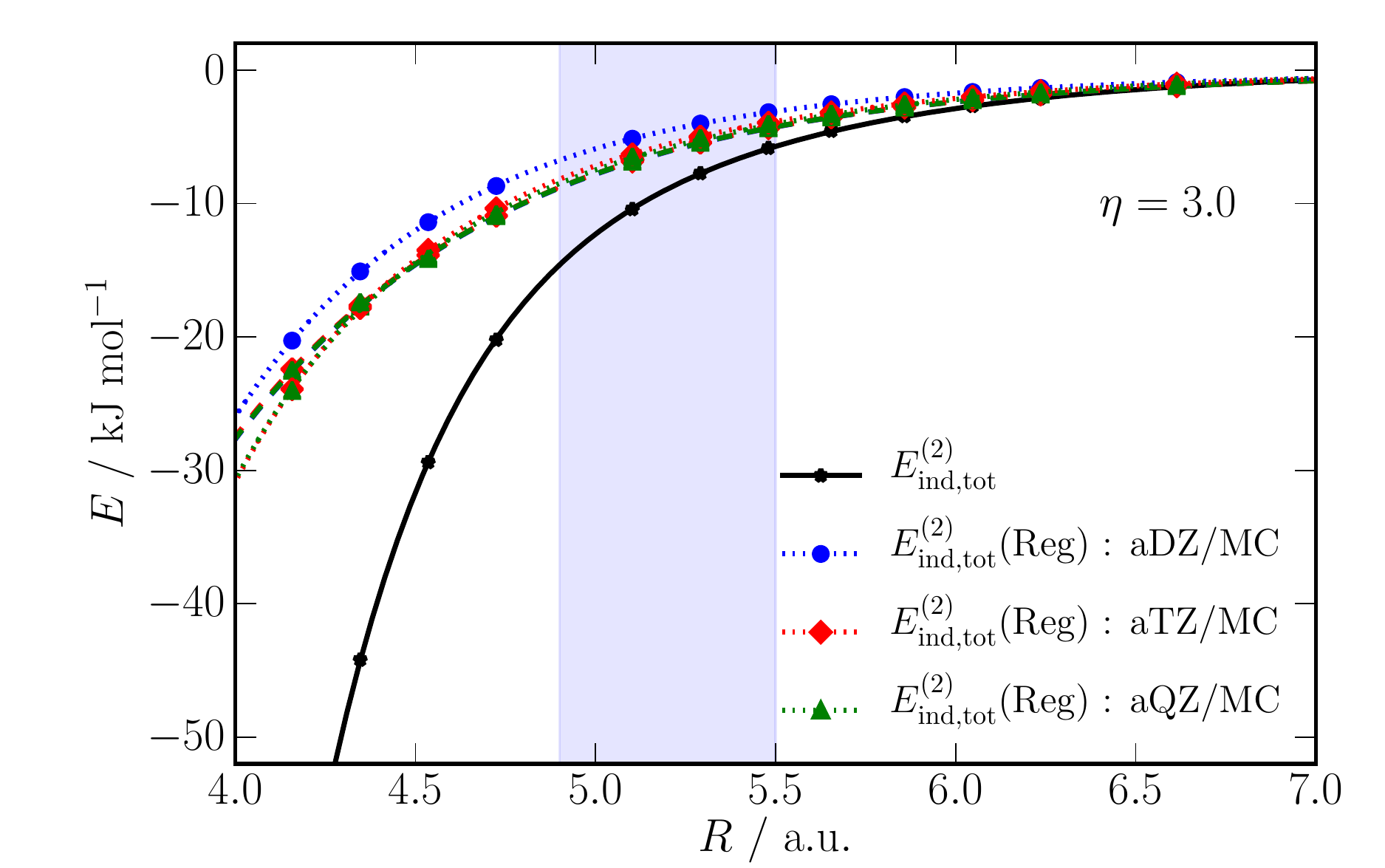}
  \includegraphics[width=0.5\textwidth]{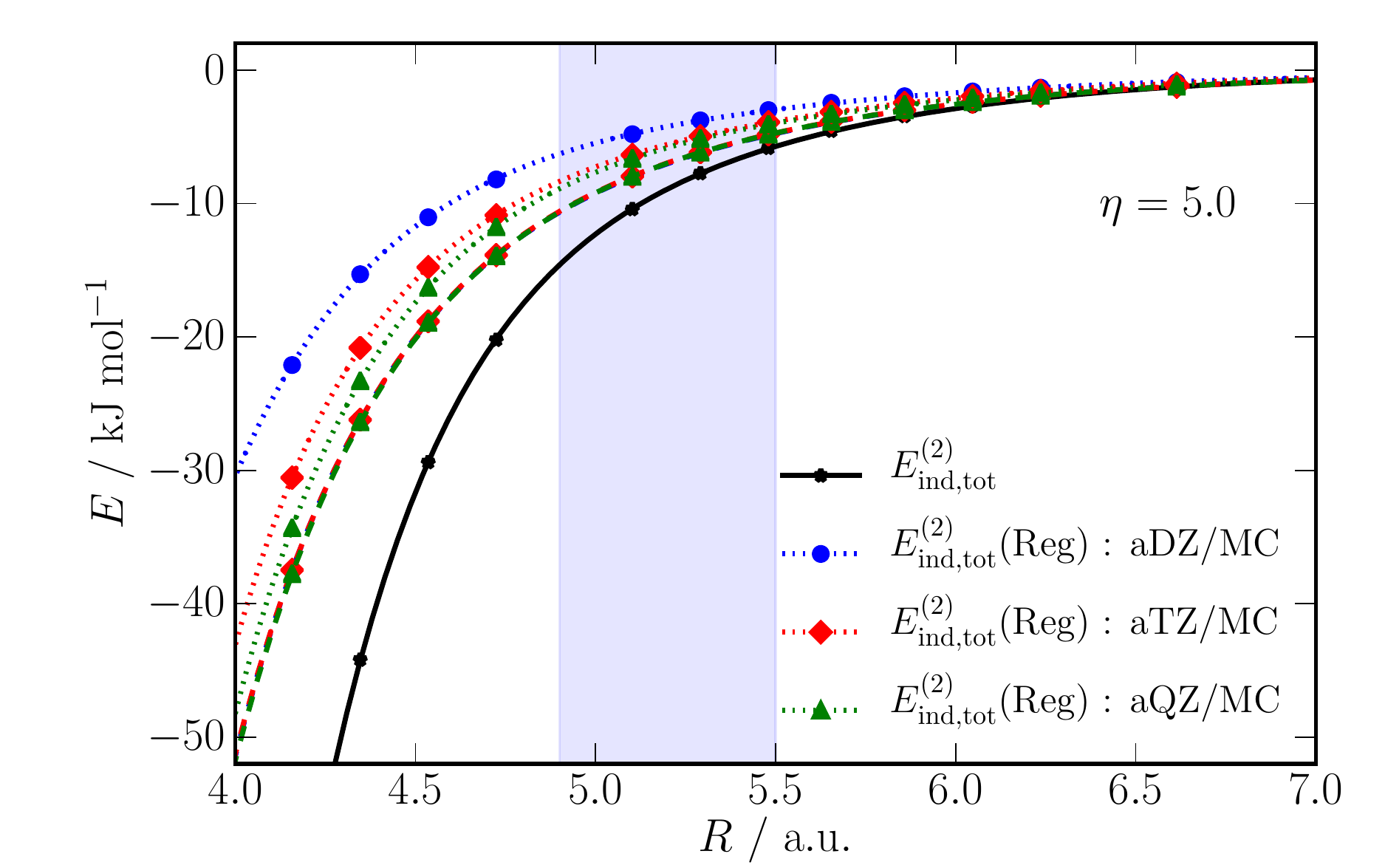}
  \caption[Water dimer in its global energy minimum configuration: 
  \Eindreg{2} energies with varying $\eta$]{
  Regularized and un-regularized total induction energies for the water dimer in its
  hydrogen-bonded orientation.
  \Eindreg{2} energies are shown for the aug-cc-pVxZ, x=D,T,Q, basis sets in the
  MC (dotted lines) and MC+ (dashed lines) types. The \Eindreg{2} energies calculated
  with the MC+ basis types are nearly indistinguishable. Only the total induction energy
  (solid black line) calculated with the aug-cc-pVQZ MC+ basis is shown.
  The shaded area is as described in the caption to fig.~\ref{fig:water2-CT-SM09-var-basis}.
  \label{fig:H2O2-min-E2indReg-vareta}
  }
\end{figure}

\begin{figure}
  \includegraphics[width=0.5\textwidth]{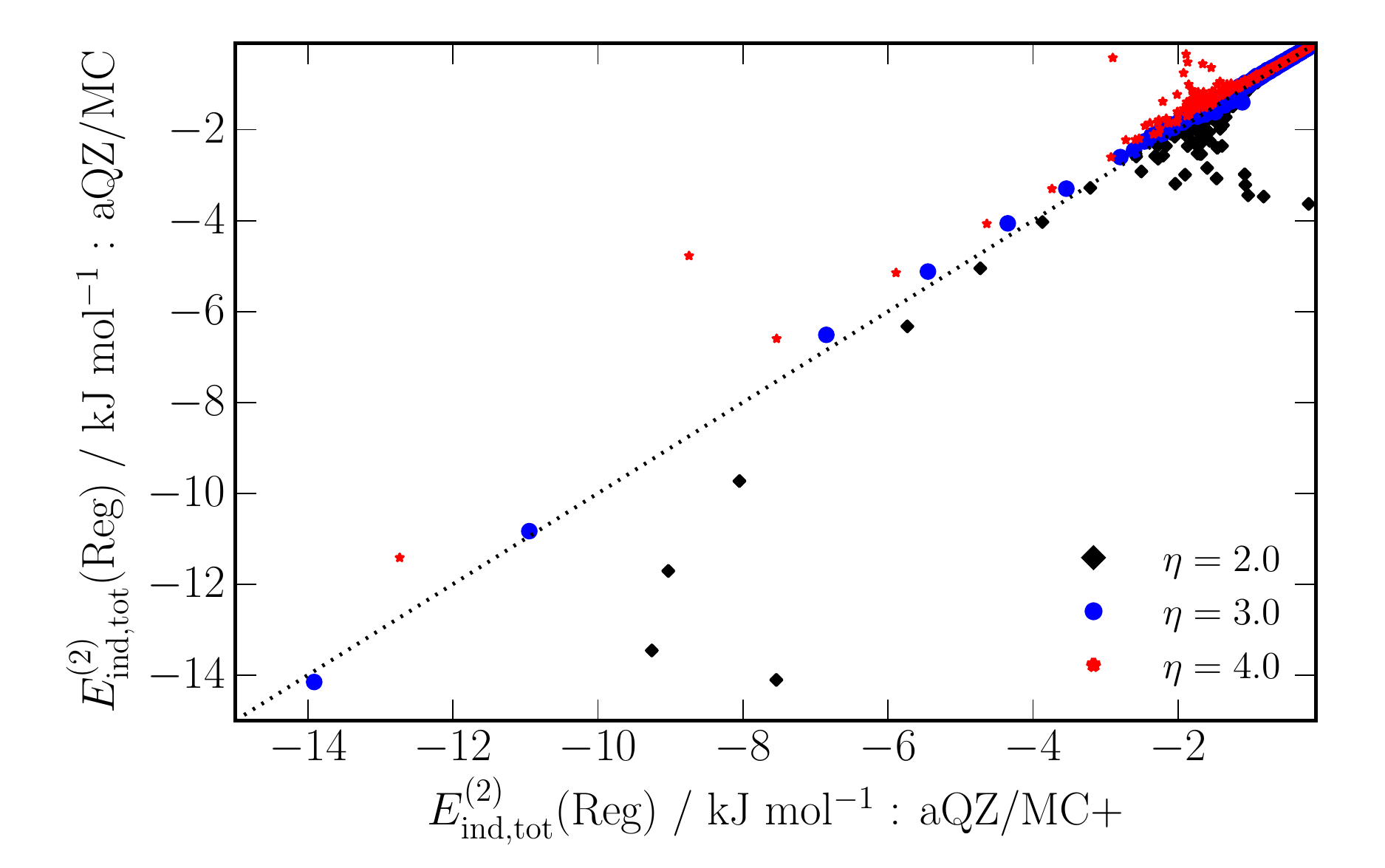}
  \caption[Water dimer in 414 geometries: \Eindreg{2} energies with varying $\eta$]{
  Second-order regularized total induction energies for the water dimer in 400 pseudo-random dimer
  configurations and 14 hydrogen-bonded configurations. Energies calculated with the 
  aug-cc-pVQZ MC basis set are plotted against those calculated with the aug-cc-pVQZ MC+
  basis set.
  \Eindreg{2} is calculated using three values of the regularization parameter $\eta$.
  The 14 H-bonded dimers have energies (with $\eta=3.0$) that are along the diagonal line.
  \label{fig:H2O2-E2indReg-vareta-scatter}
  }
\end{figure}

Consider the first proposal: Our premise here is that the true polarization
energy can be described by a reasonably large monomer basis, that is, without
the need of basis functions located on the partner monomer. Consequently,
with the correct regularization, i.e., the value of $\eta$ that completely
suppresses all charge-transfer tunneling effects without altering the
long-range part of the electrostatic potential responsible for the
polarization, all basis sets capable of describing the 
true polarization energy will yield the same, regularized, induction
energy. If on the other hand, the regularization is insufficient, then some degree
of charge-transfer will be allowed. This will lead to a spread in energies
as the larger basis sets will be better able to describe the residual tunneling.
Finally, if the regularization is excessive, the long-range part of the potential
will be affected. This will also cause a variation in energies calculated with 
different basis sets, with the larger, more diffuse basis sets being more
affected. 

In fig.~\ref{fig:H2O2-min-E2indReg-vareta} we display regularized second-order
induction energies for the water dimer calculated with various basis sets and
three values of $\eta=1.0,3.0$ and $5.0$ a.u.\
For the regularization parameter $\eta=3.0$ a.u.\ all but the smallest
basis set results in the same \Eindreg{2} energy over the entire range 
of intermolecular separations. The one exception is the aug-cc-pVDZ basis
in the monomer-centered type which is unable to adequately describe even the 
polarization component of the induction energy. For larger and smaller values
of $\eta$ the spread in the results from the six basis sets is seen to increase.
To ensure that this observation is not specific to the O$\cdots$H contact in the
water dimer, in fig.~\ref{fig:H2O2-E2indReg-vareta-scatter} we compare regularized 
induction energies for 414 water dimers: 400 of these were chosen using the pseudo-random
algorithm described in Ref.~\onlinecite{MisquittaWSP08} and the remaining
14 are those already shown in the above figures.
It is remarkable that with $\eta=3.0$ a.u.\ the regularized induction energies
calculated with the aug-cc-pVQZ MC basis and aug-cc-pVQZ MC+ basis are nearly
perfectly in agreement for all 414 dimers.
Varying $\eta$ to 2.0 or 4.0 a.u.\ results in a significantly 
less correlation between the two sets of energies, not just for the 
hydrogen-bonded dimer configurations with strong charge-transfer, but also
for configurations with O$\cdots$O contacts which exhibit small charge-transfer
energies.
We point out here that with this procedure we are unable to distinguish between
$\eta$ in the range $3.0 \pm 0.2$ a.u.\ However, it seems fairly clear that the 
optimum value is very close to $3.0$ a.u.\

\begin{figure}
  \includegraphics[width=0.5\textwidth]{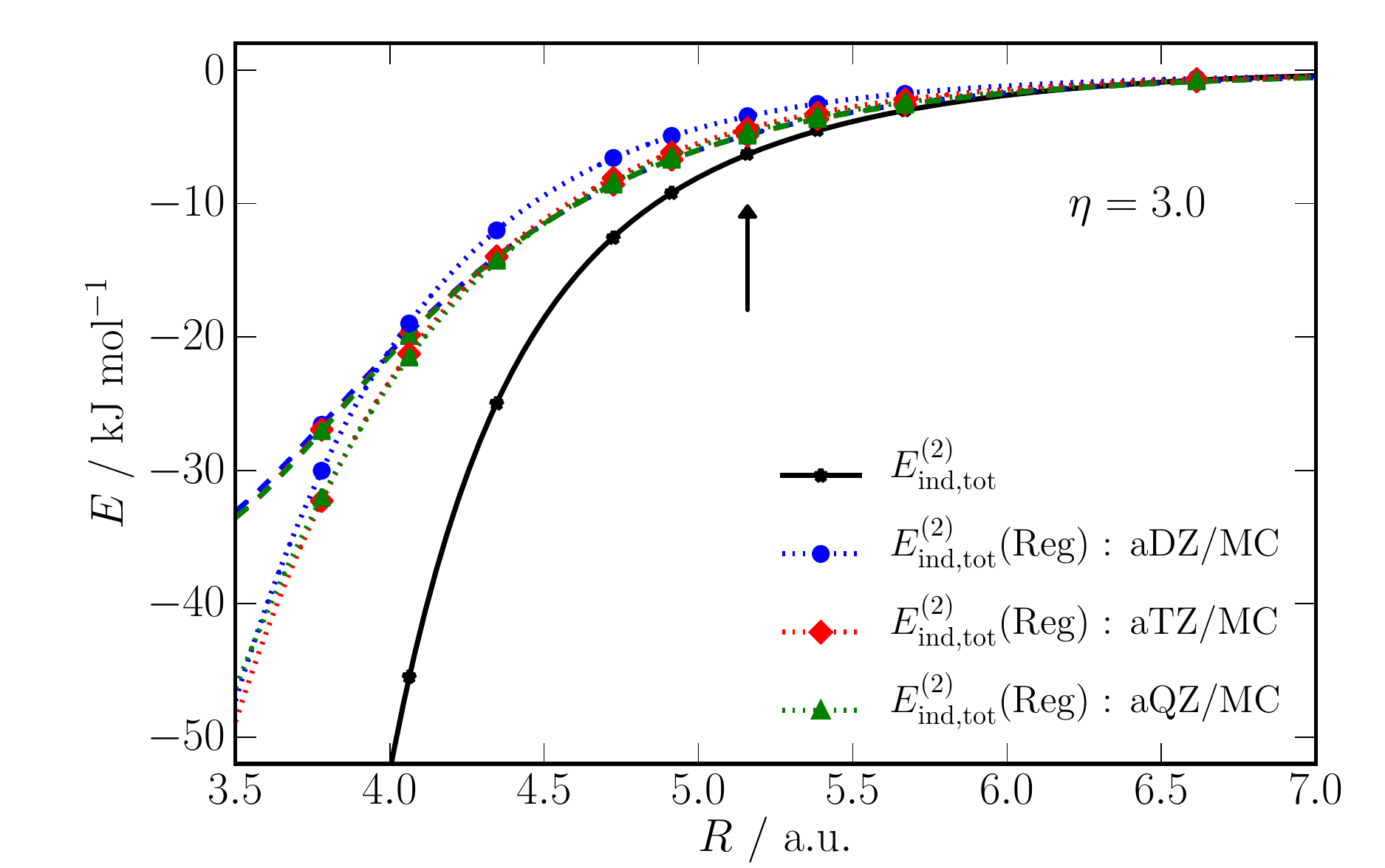}
  \caption[HF dimer in its global energy minimum configuration: 
  \CTreg{2} energies with varying $\eta$]{
  Induction and charge-transfer energies for the HF dimer at its global minimum
  hydrogen-bonded configuration. See the caption to fig.~\ref{fig:H2O2-min-E2indReg-vareta}
  for a description.
  The arrow indicates the equilibrium separation.
  \label{fig:HF2-min-vareta}
  }
\end{figure}

\begin{figure}
  \includegraphics[width=0.5\textwidth]{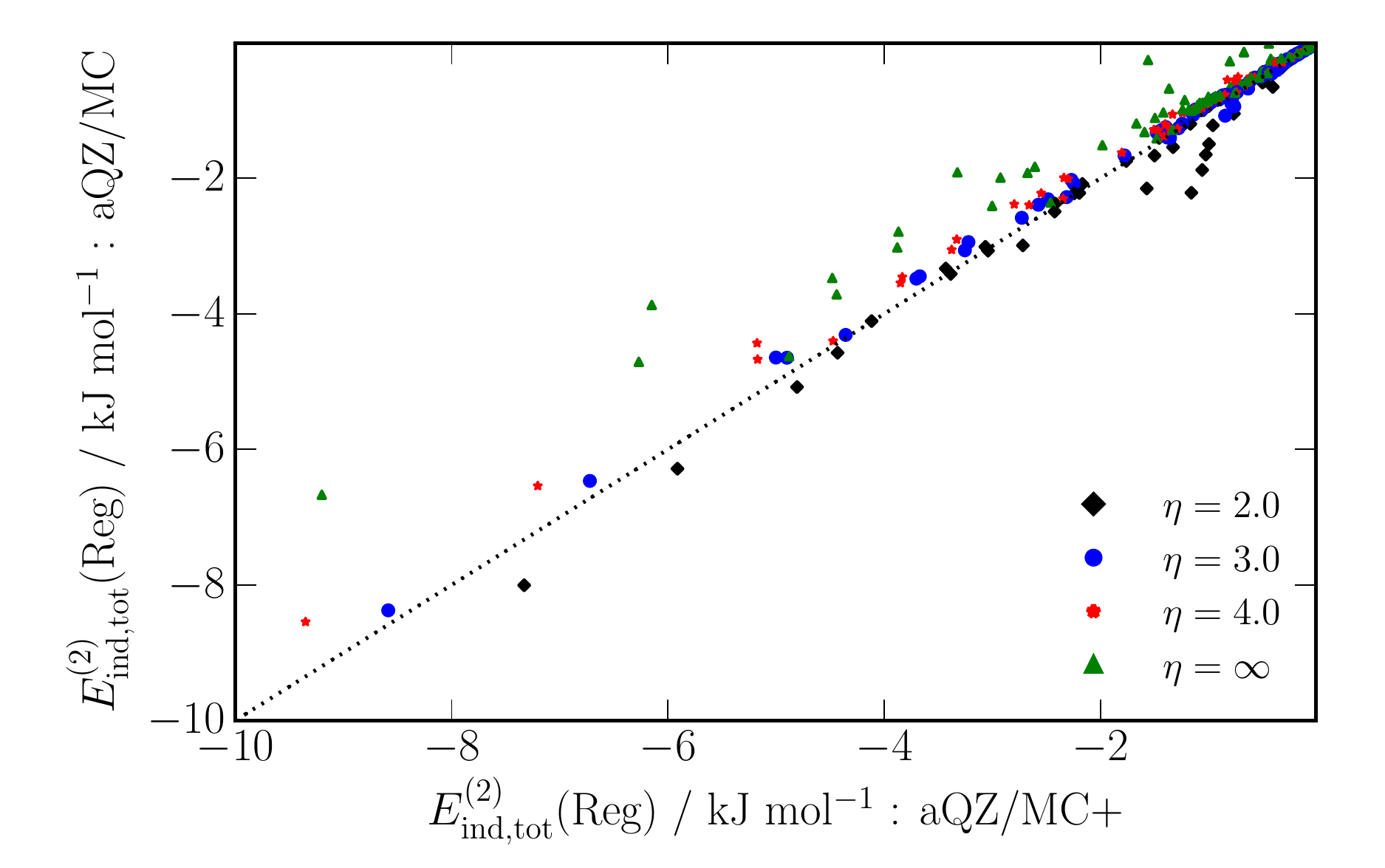}
  \caption[HF dimer in 159 geometries: \Eindreg{2} energies with varying $\eta$]{
  Second-order regularized induction energies for the HF dimer in 120 pseudo-random dimer
  configurations and three set of 13 configurations at specific geometries:
  (1) the global minimum, (2) the linear H$\cdots$H, and (3) linear O$\cdots$O
  orientations. \Eindreg{2} is calculated using three values of the
  regularization parameter $\eta$ with the aug-cc-pVQZ MC basis set values
  plotted against those calculated with the aug-cc-pVQZ MC+ basis set.
  \label{fig:HF2-E2indreg-vareta-scatter}
  }
\end{figure}

In fig.~\ref{fig:HF2-min-vareta} we display charge-transfer energies for the
HF dimer in its minimum energy hydrogen-bonded configuration \cite{PetersonD95}
calculated with various basis sets. Only results for $\eta=3.0$ a.u.\ are presented
as the behaviour of this system is largely similar to that of the water dimer: here too, 
$\eta=3.0$ a.u.\ is a good choice for the regularization parameter, though the 
regularized induction energies from the larger basis sets are not in as good agreement
at short separations at which a larger value of $\eta$ may be more appropriate. 
This may indicate that $\eta$ should depend on the type of atom. 
Although this discrepancy shows up at separations not easily accessed,
this needs further investigation.
In fig.~\ref{fig:HF2-E2indreg-vareta-scatter} we compare regularized induction energies
calculated using a few values of $\eta$ for 120 pseudo-random dimers and
39 dimers at specific orientations. We see a good agreement of regularized induction
energies calculated with the aug-cc-pVQZ MC and MC+ basis sets for a regularization
of $\eta=3.0$ a.u.\

\begin{figure}
  \includegraphics[width=0.5\textwidth]{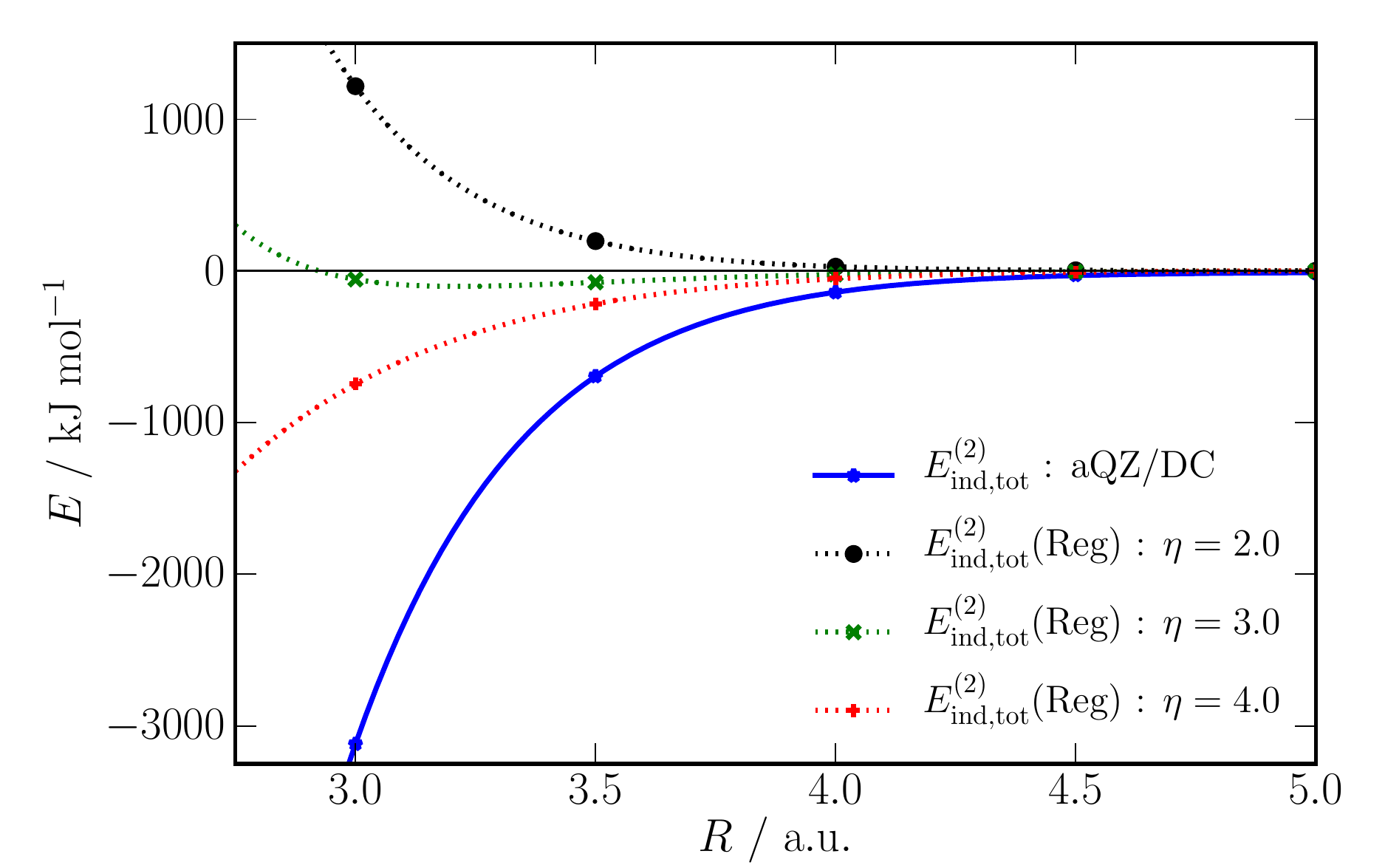}
  \caption[Ar2 regularized induction energy: varying $\eta$]{
  Argon dimer second-order induction energies without regularization and 
  regularized energies with three values of the regularization parameter $\eta$.
  All calculations used the aQZ DC-type of basis.
  These are very short inter-atomic separations: 
  the minimum energy configuration is at roughly $7.13$ a.u.\
  \label{fig:Ar2-aQZ-vareta}
  }
\end{figure}

\begin{figure}
  \includegraphics[width=0.5\textwidth]{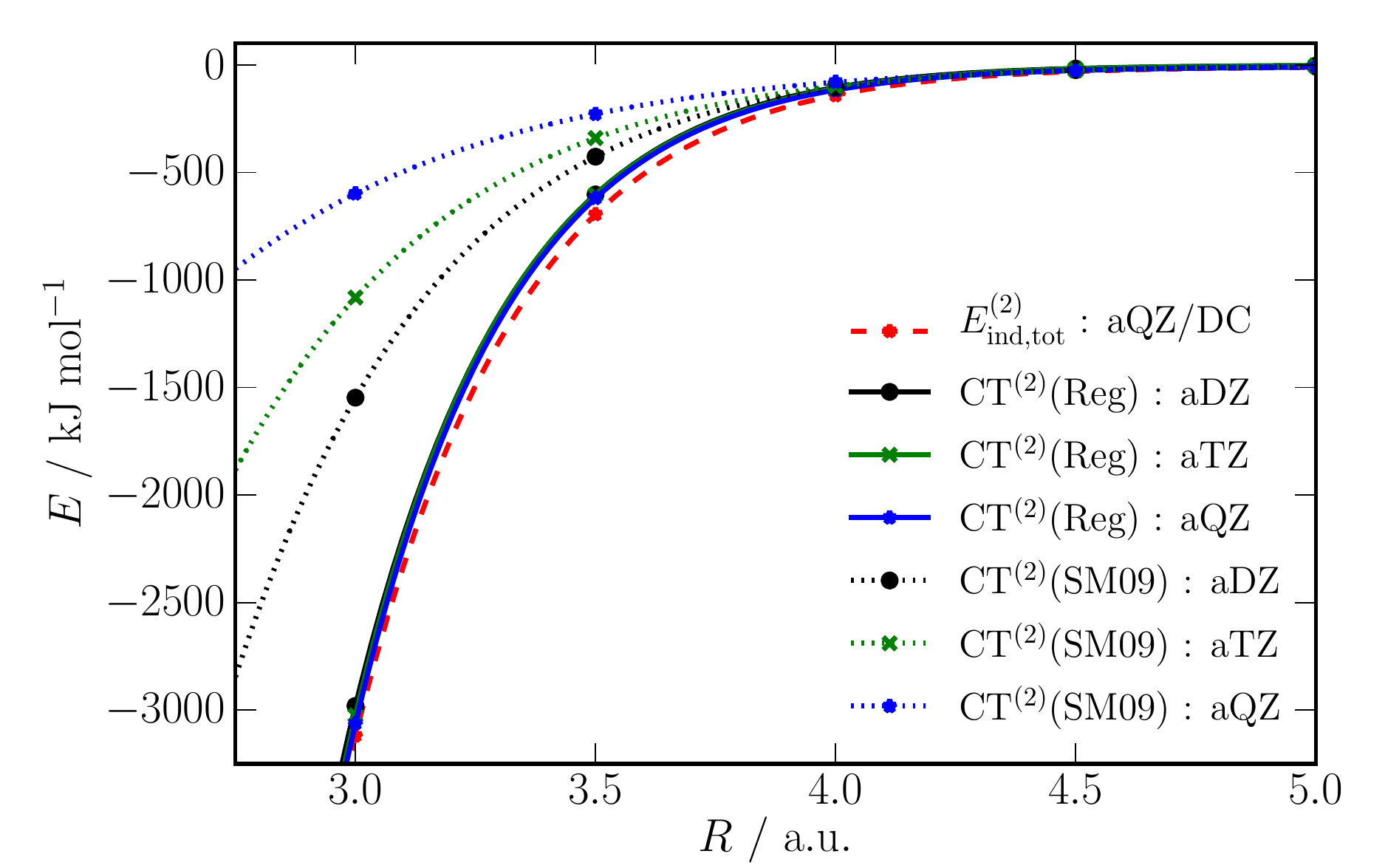}
  \caption[Ar2 : comparing CT(2)-Reg and CT(2)-SM09]{
  Argon dimer second-order induction energies (solid line) and charge-transfer energies calculated 
  using various basis sets. The second-order charge-transfer energies are calculated using 
  the regularization procedure with $\eta=3.0$ a.u.\ (dotted lines) and the 
  Stone--Misquitta procedure (dashed lines).
  \label{fig:Ar2-CT2-var-basis}
  }
\end{figure}

Now consider the second proposal: that we determine the optimum value of the
regularization parameter by requiring that all the induction energy in a
rare gas dimer is charge-transfer. There are a few issues with this 
proposal. The total induction energy of the rare gas dimers 
is nearly zero at and around the equilibrium separation, consequently
we need to use very short separations to obtain an appreciable
total induction energy. This leads to conundrum: while we can argue that
the multipole expansion should not result in any polarization energy for
these dimers, at short separations, the expansion is not valid
as the charge-densities penetrate. It is worth bearing this in mind 
in the following discussion.

The induction energy of the argon dimer, like that of
all rare-gas dimers, is almost zero at the equilibrium separation
\cite{PatkowskiMFS05} of 7.13 a.u.,
with the polarization and exchange components almost completely cancelling. 
However, as shown in fig.~\ref{fig:Ar2-aQZ-vareta}, for very small inter-atomic
separations this is no longer the case, and we get an exponentially varying
induction energy. However, with regularization parameter $\eta=3.0$ a.u.\ the
regularized induction energy is close to zero for all separations.
Using this value of $\eta$ we have calculated second-order charge-transfer
energies using the aug-cc-pV$x$Z, $x$=D,T,Q basis sets. From fig.~\ref{fig:Ar2-CT2-var-basis}
we see that \CTreg{2} is insensitive to basis set, but, as with the 
water dimer example, \CTsm{2} shows considerable basis variation, with the 
charge-transfer energy tending to zero as the basis set increases.
Unlike the first proposal, the optimum value of $\eta$ obtained in this manner
is dependent on the type of system: it is 3.2 a.u.\ for the Ar$\cdots$Ne complex,
and 2.9 a.u.\ for the Ar$\cdots$He complex. Furthermore, for the neon and
helium dimers it varies from 4.0 to 5.0 a.u.\, though, in these cases, 
the total induction energies are considerably smaller even at very short separations,
so there is a associated ambiguity in the choice of $\eta$.

Even if we set aside the second proposal, the first method provides compelling
evidence that the regularization parameter $\eta=3.0$ a.u.\ is appropriate, 
at least for the lighter atoms. This value corresponds to a regularization 
length-scale of 0.577 Bohr. 
As mentioned above, we see some indication that $\eta$ should 
vary with the type of atom, but numerical evidence suggests that this variation
is likely to be small, and is probably manifest at very short separations only.
In the rest of this article \CTreg{2} will be computed through eq.~\eqref{eq:CT-Reg}
with the regularization parameter $\eta=3.0$ a.u.

\subsection{Analysis of the second-order charge-transfer energy}
\label{ssec:analysis-of-CT2}

\begin{figure}
  \includegraphics[width=0.5\textwidth]{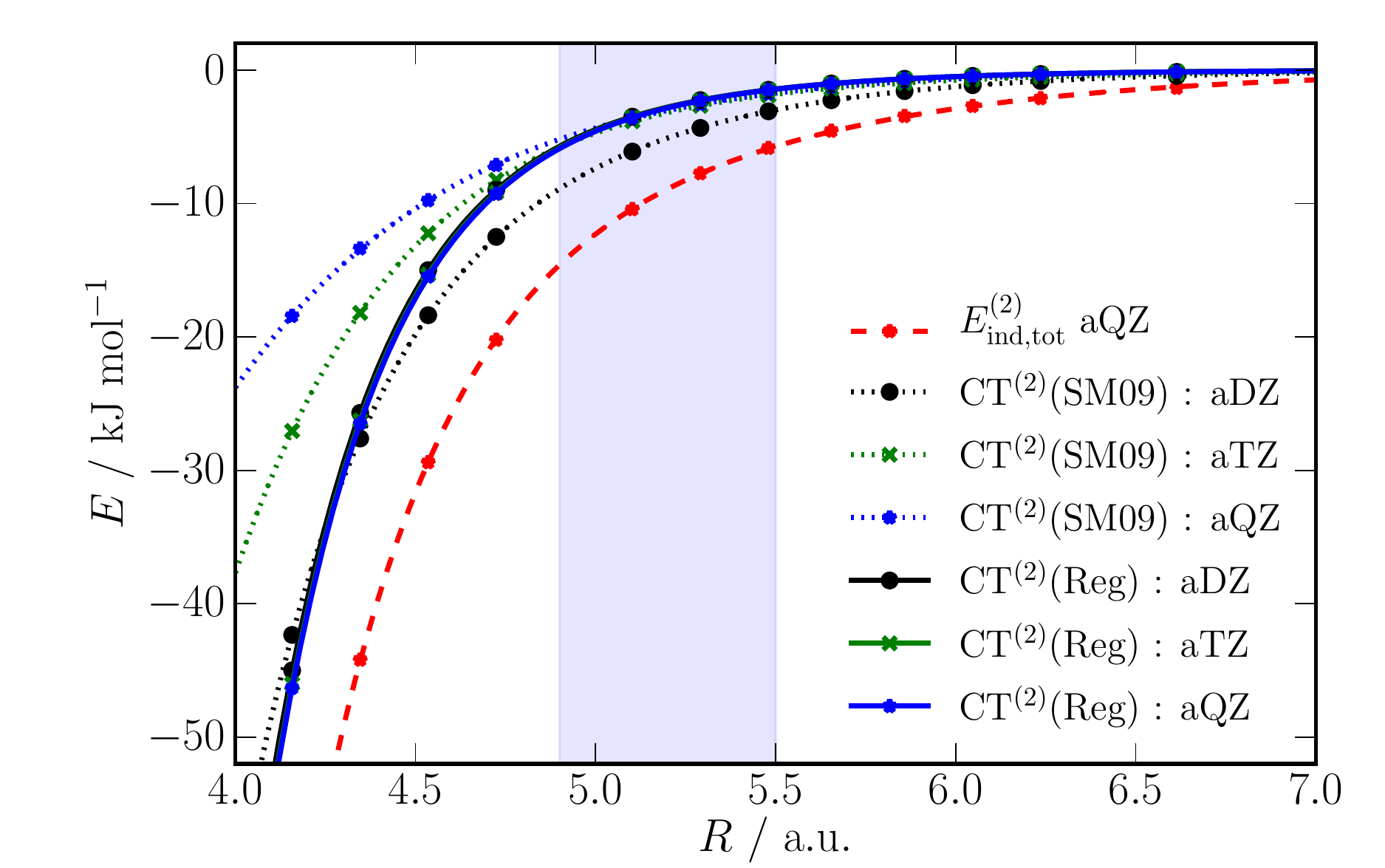}
  \caption[Water dimer in its global energy minimum configuration: 
  \CTreg{2} and \CTsm{2} energies with three basis sets]{
  Second-order charge-transfer energies for the water dimer in its hydrogen-bonded
  orientation.
  \CTreg{2} (solid lines) and \CTsm{2} (dotted lines) energies are shown for the
  aug-cc-pVxZ, x=D,T,Q, basis sets. All \CTreg{2} calculations were performed
  with the MC+ basis type and show virtually no variation with basis.
  The total (un-regularized) second-order induction energy
  (solid black line) calculated with the aug-cc-pVQZ MC+ basis is shown.
  The shaded area is as described in the caption to fig.~\ref{fig:water2-CT-SM09-var-basis}.
  \label{fig:H2O2-min-CT2-var-basis}
  }
\end{figure}

Having determined the manner in which the regularization is to be performed,
we will now analyse the \CTreg{2} energies in detail. 
In fig.~\ref{fig:H2O2-min-CT2-var-basis} we display \CTreg{2} and \CTsm{2} energies
for the water dimer. The \CTreg{2} energies have been calculated using \eqref{eq:CT-Reg}
with both terms on the RHS calculated using the MC+ type of basis and therefore show
very little sensitivity to the choice of basis set (as long as augmented double-$\zeta$ or
better in quality). Contrast this with the strong basis sensitivity of the \CTsm{2} 
energies. There are a few features of these energies worth highlighting: at long-range, 
the \CTreg{2} energies are similar to those of \CTsm{2} with the aug-cc-pVQZ basis set.
But at short-range, \CTreg{2} is closer to \CTsm{2} with the much smaller aug-cc-pVDZ basis set.
This behaviour supports the discussion in Sec.\ \ref{ssec:CT-basis-space-defns}: 
The \CTsm{2} definition does result in accurate second-order charge-transfer energies
with the aug-cc-pVQZ basis set at long-range, but as the intermolecular separation 
decreases, this large basis set starts to pick up charge-transfer excitations leading
to an underestimation of the charge-transfer defined via eq.~\eqref{eq:CT-SM09}.
At short separations, we should expect a smaller basis set, in this case, the 
aug-cc-pVDZ basis, to yield more accurate \CTsm{2} energies. 
This is indeed what our \CTreg{2} definition demonstrates.

\begin{figure}

  \includegraphics[width=0.5\textwidth]{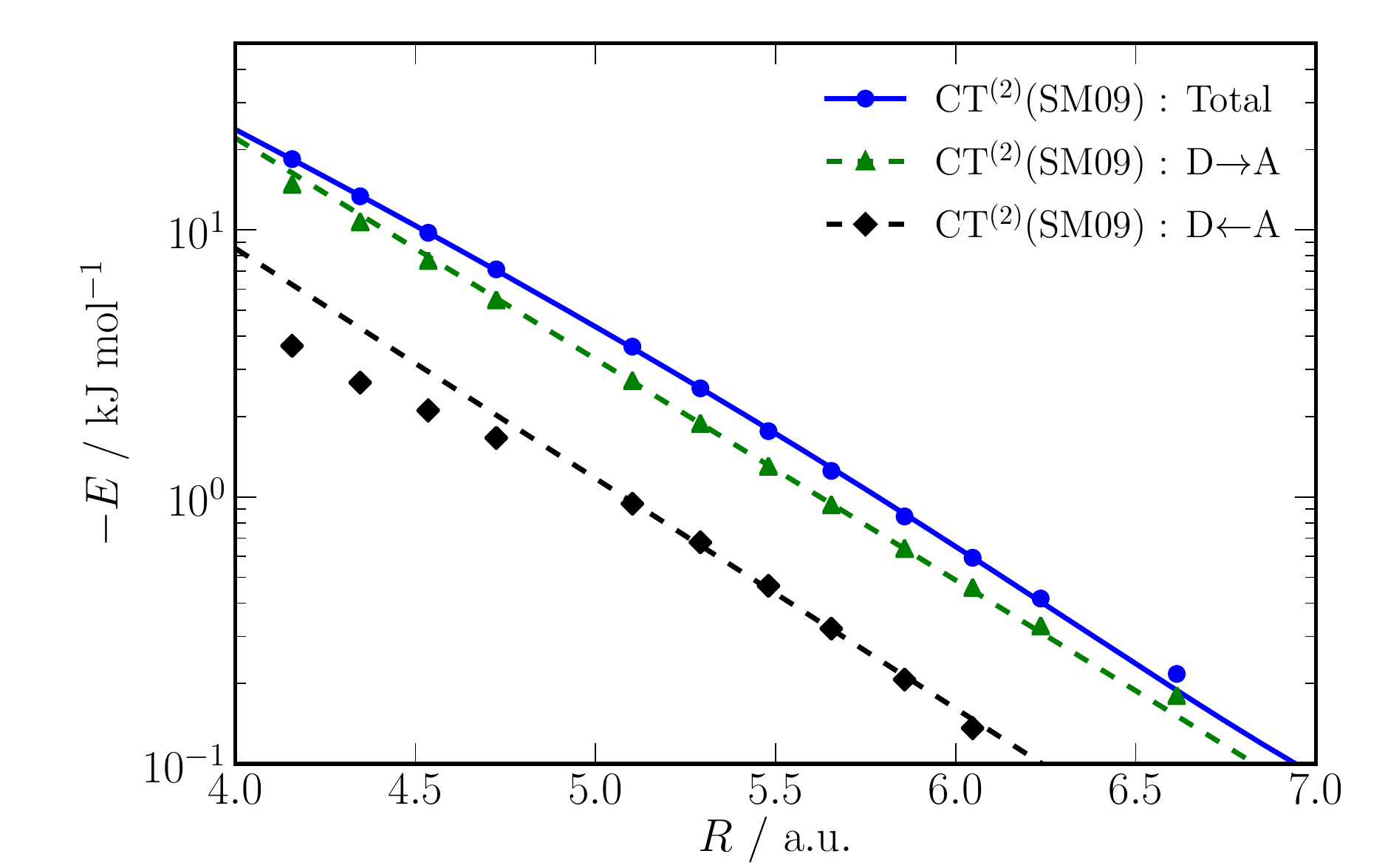}
  \includegraphics[width=0.5\textwidth]{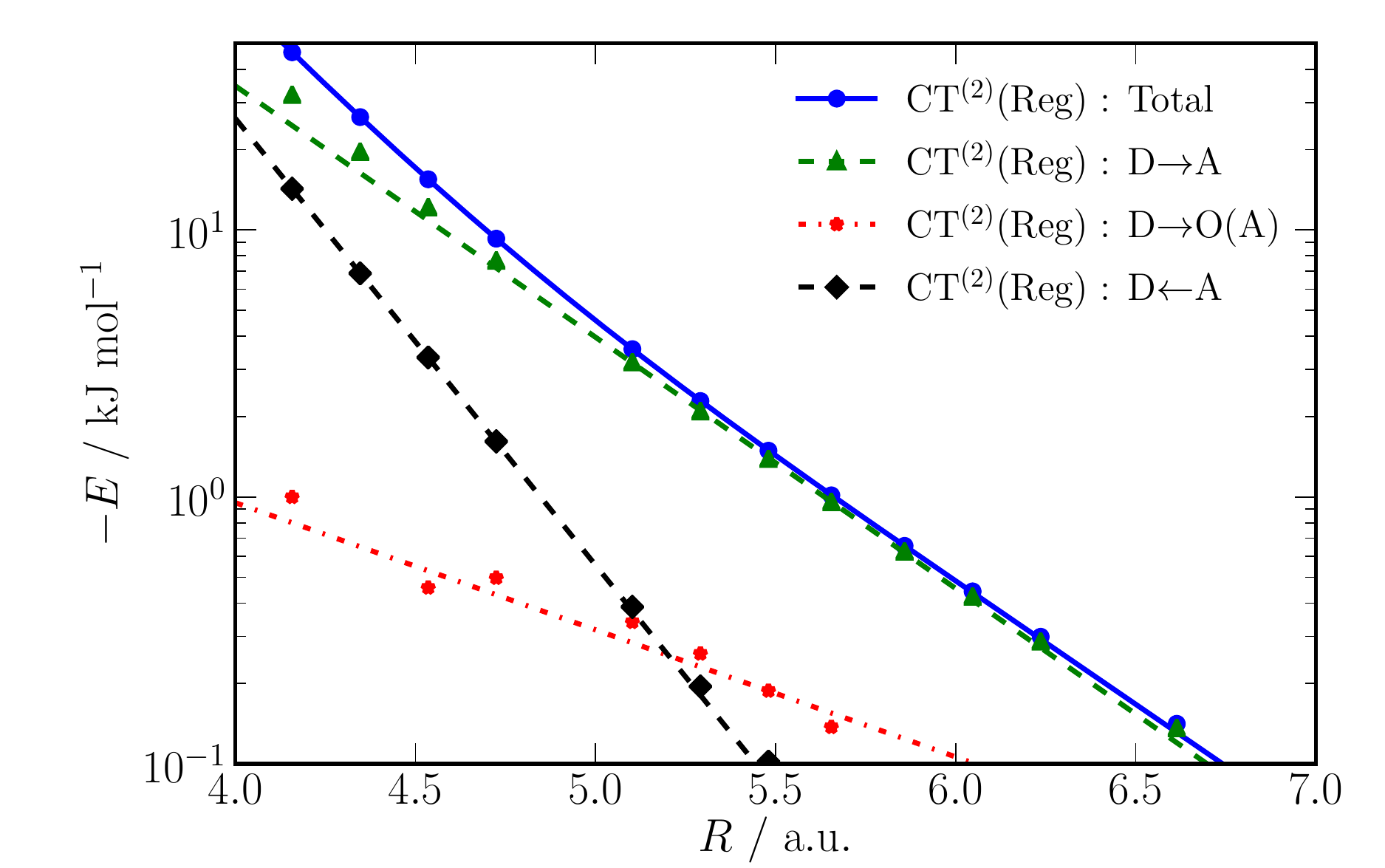}
  \caption[Water dimer in its global energy minimum configuration: 
  semilog plots of \CTreg{2} and \CTsm{2} energies]{
  Decomposition of the second-order charge-transfer energies for the water dimer
  in its hydrogen-bonded orientation. All calculations used the aug-cc-pVQZ basis set.
  The second-order charge-transfer (solid blue lines) can be decomposed into 
  a contribution from the electron donor (oxygen) to the electron acceptor (hydrogen) (green, dashed) 
  and another, weaker contribution, from the acceptor (hydrogen) to the donor (oxygen)
  (black, dashed). Furthermore, the regularization procedure
  allows us to isolate a contribution from the donor to the oxygen of the 
  acceptor (red, dotted).
  \label{fig:H2O2-min-CT2-details}
  }
\end{figure}

In fig.~\ref{fig:H2O2-min-CT2-details} we present the charge-transfer energies calculated in
the aug-cc-pVQZ basis set on a semi-log scale. The charge-transfer is usually thought of as
being decaying exponentially with separation, so on this plot it should appear as a 
straight line. However, this is not the case for either of the methods.
The \CTsm{2} energy does exhibit an exponential behaviour at large intermolecular
separations, but becomes sub-exponential at short separations. This is a consequence of the
problem discussed in the above paragraph: the \CTsm{2} will always result in too 
little CT at short range, and this effect is larger for the larger basis sets.
By contrast, \CTreg{2} is super-exponential at short separations: it exhibits 
a clear double (possibly even multiple) exponential behaviour. This should be expected.
If we accept that the charge-transfer process is a tunneling phenomenon, then we should
also expect to see contributions from tunneling into each of the (screened) nuclear wells.
The dominant process is expected to be the charge-density of the electron donor (oxygen)
tunneling into the poorly screened nuclear potential of the electron acceptor (hydrogen).
(In this paper, we use the terms `donor' and `acceptor' to refer to electrons and not
protons.)  There will also be 
tunneling of the acceptor density into the well-screened nuclear potential of the
donor oxygen. However, we may additionally expect weaker processes such as
the donor density tunneling into the acceptor oxygen. Each process will be approximately
exponential, with the barrier height and width determining the exponent.
Perturbation theory allows us make the donor to acceptor and acceptor to donor decomposition.
These results are displayed in fig.~\ref{fig:H2O2-min-CT2-details}.
The decomposition of \CTsm{2} is qualitatively different from that of \CTreg{2}. In the
former we see two processes both apparently with the same exponent (except at short-range).
These do not reflect the tunneling processes described above.
However, \CTreg{2} exhibits at least two exponential processes:
The acceptor to donor energy has a larger exponent and decays rapidly with separation. 
This is what we would expect as tunneling into the well-screened donor oxygen potential
must proceed through a large barrier, leading to a large exponent.

The regularization procedure allows us to analyse the charge-transfer process
in even more detail as we can selectively regularize the nuclear potential of
the acceptor hydrogen atoms and isolate the charge-transfer contribution from
the donor to the oxygen of the acceptor. 
From fig.~\ref{fig:H2O2-min-CT2-details}  we see that, as might be expected,
this is a much weaker process. What is unusual is the relatively small exponent
associated with this process. The reason for this is as yet unclear.

\begin{figure}
  \includegraphics[width=0.5\textwidth]{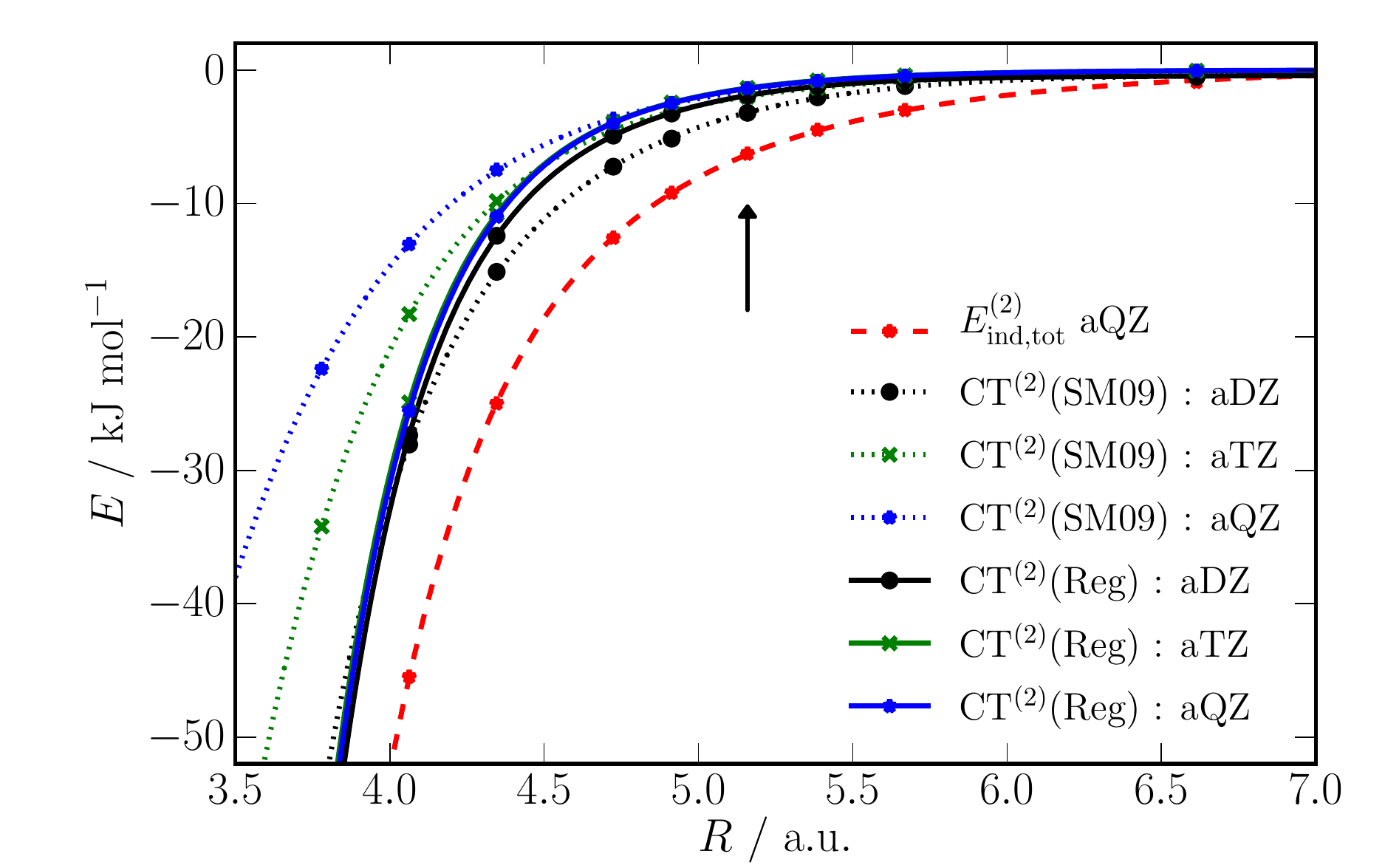}
  \caption[HF dimer in its global energy minimum configuration: 
  \CTreg{2} and \CTsm{2} energies with three basis sets]{
  Second-order charge-transfer energies for the HF dimer in its hydrogen-bonded
  orientation.
  See the caption to fig.~\ref{fig:H2O2-min-CT2-var-basis} for a description.
  The arrow indicates the equilibrium separation.
  \label{fig:HF2-min-CT2-var-basis}
  }
\end{figure}

\begin{figure}

  \includegraphics[width=0.5\textwidth]{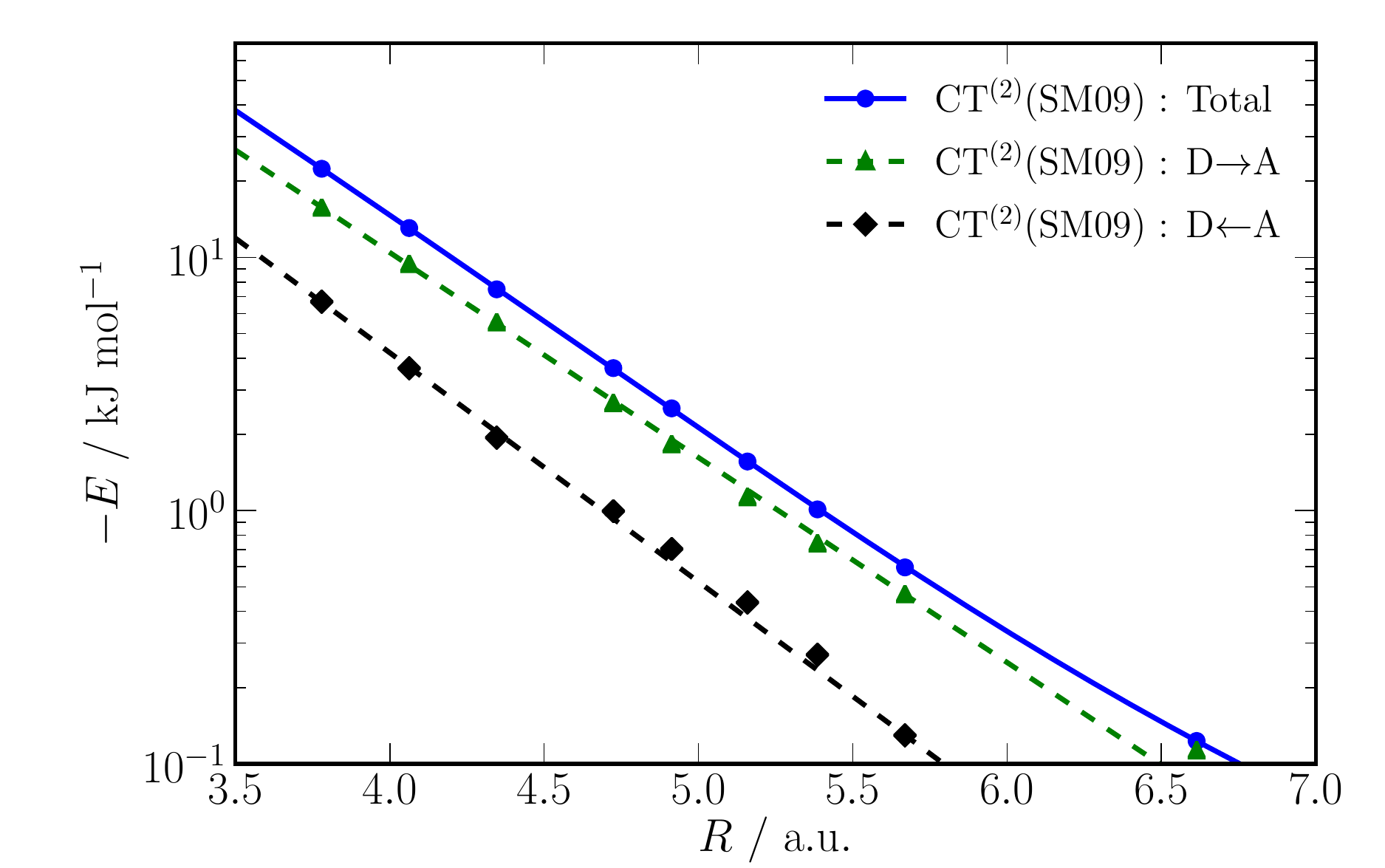}
  \includegraphics[width=0.5\textwidth]{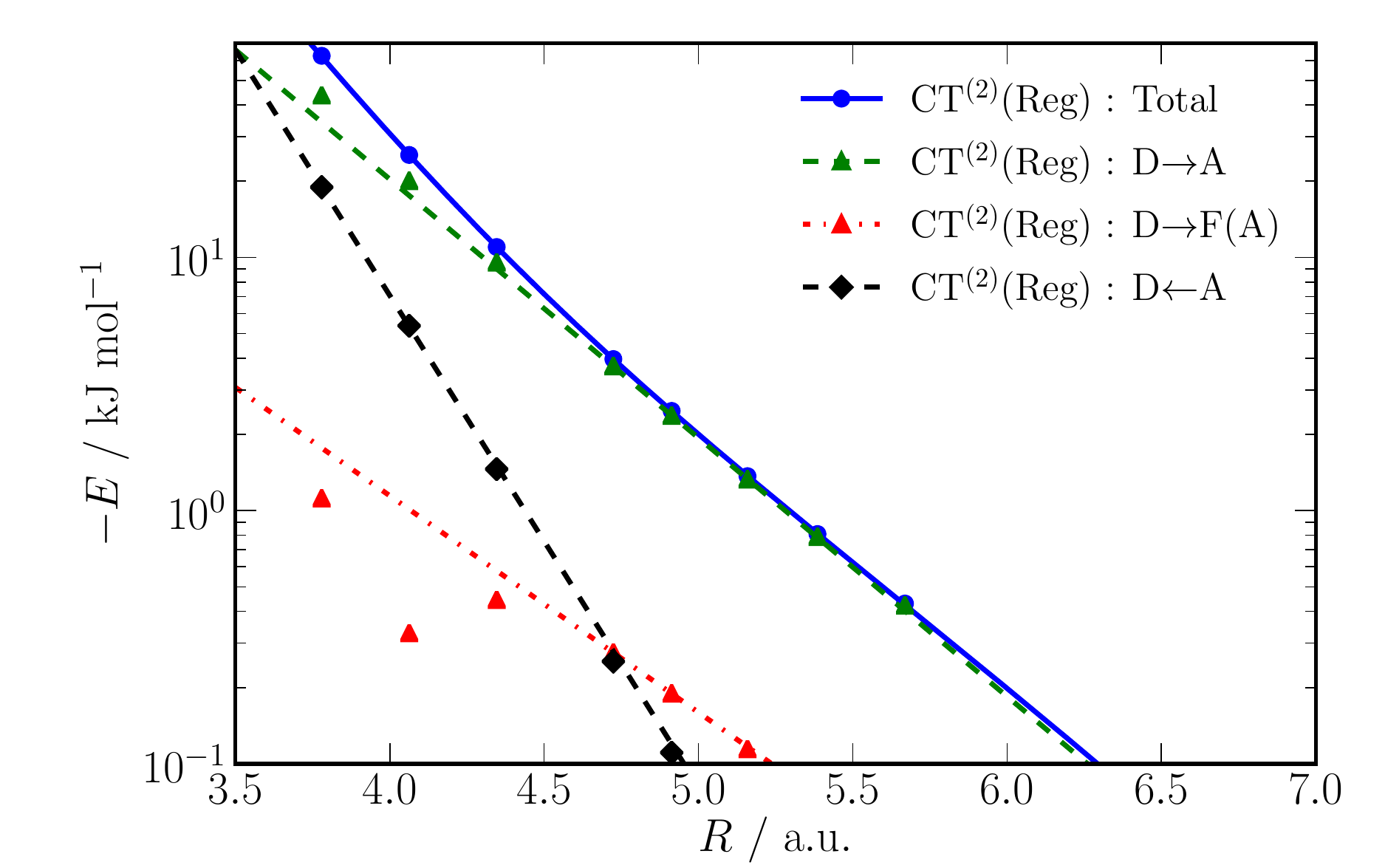}
  \caption[HF dimer in its global energy minimum configuration: 
  semilog plots of \CTreg{2} and \CTsm{2} energies]{
  Decomposition of the second-order charge-transfer energies for the HF dimer
  in its hydrogen-bonded orientation. 
  See the caption to fig.~\ref{fig:H2O2-min-CT2-details} for a description.
  \label{fig:HF2-min-CT2-details}
  }
\end{figure}

In figs.~\ref{fig:HF2-min-CT2-var-basis} and \ref{fig:HF2-min-CT2-details}
we present a similar analysis of the second-order charge-transfer energy for the
HF dimer in its hydrogen-bonded orientation. The features we have highlighted
for the water dimer are also seen here. As with the water dimer, \CTreg{2}
is largely basis-independent (the aug-cc-pVTZ/MC+ and aug-cc-pVQZ/MC+ results
are nearly identical, but the aug-cc-pVDZ/MC+ values differ) and
interpolates between the \CTsm{2} values calculated using the
aug-cc-pVDZ basis (at short-range) and the aug-cc-pVQZ basis (at long-range).
Also, as for the water dimer, the acceptor to donor second-order charge-transfer
energy becomes appreciable only for short dimer separations (less than 5 a.u.).
This is quite different from the \CTsm{2} energy for which we see 
acceptor to donor contributions at all separations.

\begin{table*}[Ht]
\setlength{\tabcolsep}{7pt}
\caption[CT-Tab1]{
Second-order induction energies and charge-transfer energies for several systems 
at specified geometries. Unless otherwise stated, all calculations use the 
aug-cc-pVQZ basis set. The (electron) donor to (electron) acceptor (D$\rightarrow$A) and 
acceptor to donor (D$\leftarrow$A) contributions are shown.
For the water dimer, the longer $R_{\rm OH}$ distance of 3.67 a.u.\ is 
representative of the O$\cdots$H separation in the dimer and the shorter distance
of 3.29 a.u.\ is representative of the separation in clusters of water molecules.
The HF dimer is at its equilibrium geometry, while the HF$\cdots$CO dimers
are both linear structures close to their radial equilibria.
Both H$_3$B structures are optimized, and the pyridine dimer is in its doubly
hydrogen-bonded planar, $D_{\rm 2h}$ symmetry configuration.
The results from Khaliullin \etal (Ref.\onlinecite{KhaliullinBH-G08}) are
infinite-order charge-transfer energies calculated using the 
$6-31(2+,2+)\mbox{G}(df,pd)$ basis set.
All energies are reported in \kJmol and distances in Bohr.}
\label{tab:CT-various-systems} 
\begin{center}
\begin{tabular}{ l *{8}{C}}
\toprule
 $R$/Bohr& \multicolumn{2}{c}{\Eind{2}}
                              & \multicolumn{2}{c}{\CTreg{2}}
                                                     & \multicolumn{2}{c}{\CTsm{2}}
                                                                          & \multicolumn{2}{c}{Khaliullin \etal
                                                                                   Ref.\onlinecite{KhaliullinBH-G08}} \\
         & \mbox{D}\rightarrow\mbox{A}
                   & \mbox{D}\leftarrow\mbox{A}
                              & \mbox{D}\rightarrow\mbox{A}
                                          & \mbox{D}\leftarrow\mbox{A}
                                                    & \mbox{D}\rightarrow\mbox{A}
                                                               & \mbox{D}\leftarrow\mbox{A}
                                                                          & \mbox{D}\rightarrow\mbox{A}
                                                                                      & \mbox{D}\leftarrow\mbox{A} \\
\midrule
\multicolumn{9}{l}{Water dimer in hydrogen-bonded orientation} \\
$R_{\rm OH} = 3.67$  
         &  -4.59  &  -1.25   &  -1.39   &  -0.10   &  -1.30   &  -0.46   &\mbox{---} &\mbox{---} \\
$R_{\rm OH} = 3.29$  
         &  -8.64  &  -1.79   &  -3.19   &  -0.39   &  -2.71   &  -0.94   &\mbox{---} &\mbox{---} \\[6pt]
\multicolumn{9}{l}{HF dimer in hydrogen-bonded orientation} \\
$R_{\rm FH} = 3.44$  
         &  -5.90  &  -0.37   &  -1.33   &  -0.04   &  -1.13   &  -0.43   &\mbox{---} &\mbox{---} \\[6pt]       
\multicolumn{9}{l}{FH$\cdots$CO (linear)} \\
$R_{\rm HC} = 4.0$
         &  -6.24  &  -0.23   &  -1.39   &  -0.05   &  -1.28   &  -0.17   &\mbox{---} &\mbox{---} \\[6pt]       
\multicolumn{9}{l}{FH$\cdots$OC (linear)} \\
$R_{\rm HO} = 4.0$
         &  -3.04  &  -0.02   &  -0.37   &  -0.02   &  -0.37   &  -0.13   &\mbox{---} &\mbox{---} \\[6pt]       
\multicolumn{9}{l}{H$_3$B$\cdots$CO : aug-cc-PVTZ/DC : B3LYP optimized (linear)} \\
$R_{\rm BC} = 2.89$
         & -296.24 &  -50.45  & -139.16  & -31.85   & -20.16   & -11.73   &  -123     &  -128     \\[6pt]       
\multicolumn{9}{l}{H$_3$B$\cdots$NH$_3$ : aug-cc-PVTZ/DC : B3LYP optimized} \\
$R_{\rm BN} = 3.21$
         & -160.77 &  -14.83  & -61.65   & -9.84    & -14.86   &  -7.23   &  -130     &  -11      \\[6pt]       
\multicolumn{9}{l}{Pyridine dimer: aug-cc-PVTZ/MC+ : $D_{\rm 2h}$ double H-bonded geometry} \\
$R_{\rm NH} = 4.82$
         &   -1.63 &  -1.63   &  -0.18   &  -0.18   &  -0.41   &  -0.41   &\mbox{---} &\mbox{---} \\[6pt]
\bottomrule
\end{tabular}
\end{center}
\end{table*}

Table\ \ref{tab:CT-various-systems} shows numerical values of the 
second-order induction and charge-transfer energies for various dimers.
The results for the water and HF dimers quantify what is already 
displayed in the above figures: at the selected geometries, \CTreg{2} and 
\CTsm{2} agree reasonably well; but the differences get larger at shorter separations.
With the exception of the H$_3$B complexes, the charge-transfer is mainly
from the donor to the acceptor; the reverse process (acceptor to donor) 
is much weaker. This is particularly true for \CTreg{2}; although
the donor to acceptor energy is dominant for \CTsm{2}, the acceptor to donor 
energies can be significantly larger than those from \CTreg{2}.
For the mixed HF and CO system, charge-transfer from the C to H in 
FH$\cdots$CO is nearly 3 times as large as that from O to H in 
FH$\cdots$OC, consistent with the strong $\sigma$-donor property of
C in CO.

The H$_3$B complexes with CO and NH$_3$ are interesting as both of these 
have very short separations. These separations are short enough that one 
may question the use of an intermolecular perturbation theory like 
SAPT(DFT). Perhaps surprisingly, SAPT(DFT) interaction energies
are within 5\% of CCSD(T) energies for both complexes, with the agreement
between the two best for the H$_3$B$\cdots$NH$_3$ complex.
The differences between \CTreg{2} and \CTsm{2} are quite large for both 
complexes. As discussed in sec.\ \ref{ssec:CT-basis-space-defns}, 
the CT-SM09 definition should be expected to recover an ever smaller 
fraction of the true second-order CT as the intermolecular separation 
decreases. This seems to be the case here. While \CTsm{2} correctly 
identifies the donor (NH$_3$, CO) to acceptor (H$_3$B) charge-transfer as 
the larger quantity, the actual amount of CT from this method
is substantially smaller than both \CTreg{2} and the infinite-order
results of Khaliullin \etal \cite{KhaliullinBH-G08}. 
Interestingly, while \CTreg{2} is consistent with the Khaliullin results
for the H$_3$B$\cdots$NH$_3$ complex, the two sets of results differ qualitatively
for the H$_3$B$\cdots$CO complex. Here we would expect transfer from 
the CO to the H$_3$B to dominate. This is given by \CTreg{2}, but 
the ALMO method shows the opposite result. 

Finally, the pyridine dimer is an interesting case as it exhibits a 
double hyhrogen bond between each of the N$\cdots$H pairs in the 
$D_{\rm 2h}$ symmetry, planar complex. The donor (N) to acceptor (H)
bond length is the longest of the complexes discussed above.
This leads to relatively small total induction energies and 
even smaller charge-transfer energies. The latter are just
over a tenth of the total induction energy---nearly an order
of magnitude smaller than the charge-transfer energies in the other
complexes. In this case the electron delocalisation process works
symmetrically in both directions, consequently there is no nett 
charge transfered between the two pyridine molecules. 

\subsection{Polarization models}
\label{ssec:pol-models}

Now that we have demonstrated both the numerical stability and the physical
nature of \CTreg{2}, we are in a position to use this definition to determine
the damping needed in polarization models. The basic idea is to fit the 
damping parameters so that the (non-iterated) classical polarization model
matches the true second-order polarization energies for a large number of
dimers. In the following examples the classical polarization model was
constructed using a rank 3 (3 on the heavy atoms and 1 on the 
hydrogen atoms) WSM distributed polarizability model \cite{MisquittaS08a,MisquittaSP08}
and a rank 4 (all atoms) distributed multipole model \cite{Stone05}.
The models are damped using the Tang--Toennies damping functions 
\cite{TangT84} with isotropic (possibly site-pair-dependent) damping parameters.
For numerical details see Sec.\ \ref{ssec:numerical}.
Note that the damping coefficients obtained in this section depend 
on the details of the model, consequently a direct comparison to other
attempts to determine the damping coefficients \cite{SebetciB10a} cannot be 
made.

\begin{figure}
  \includegraphics[width=0.5\textwidth]{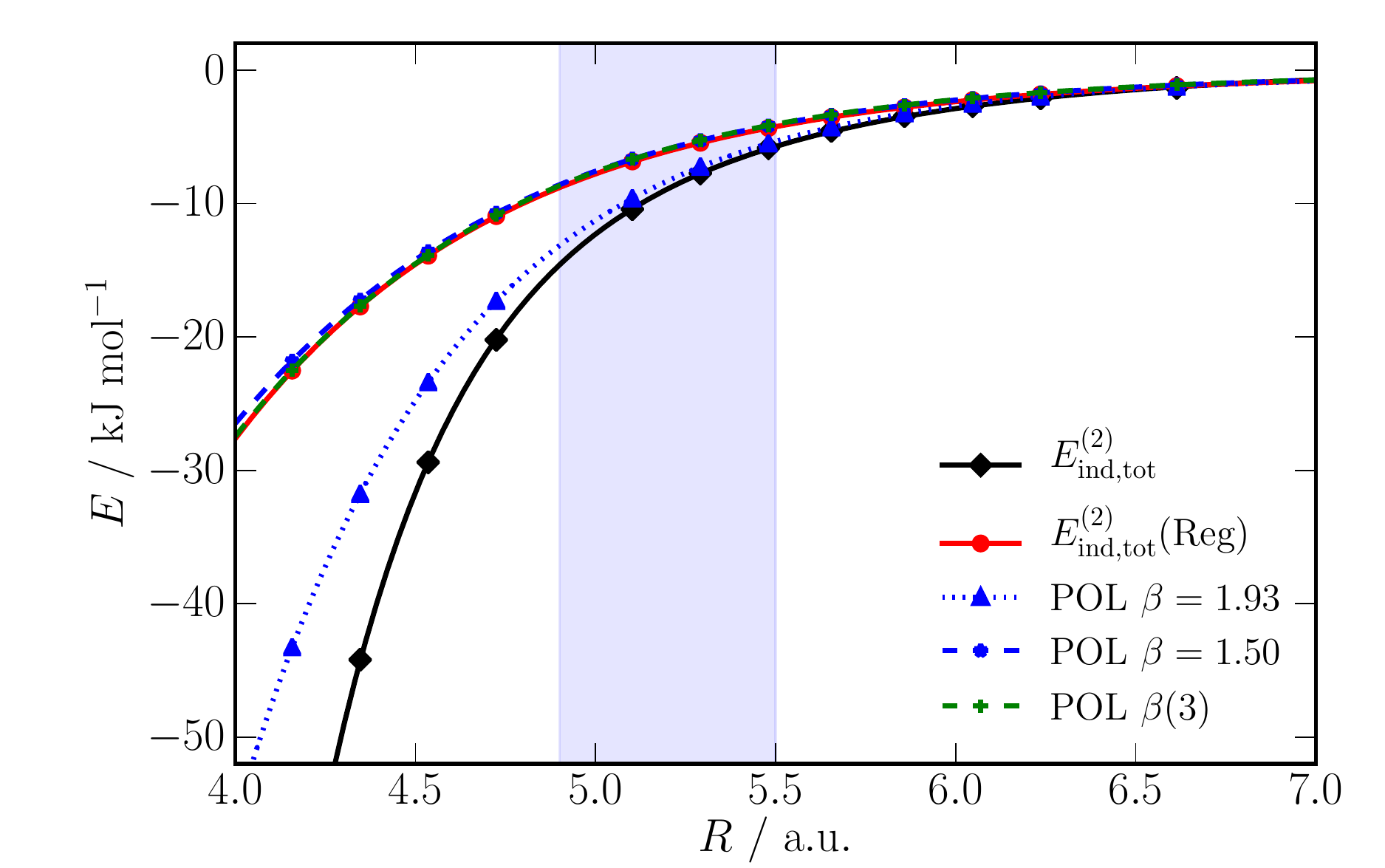}
  \caption[Water dimer in its global energy minimum configuration: 
  polarization models]{
  Second-order polarization energies for the water dimer in its hydrogen-bonded
  orientation. Reference polarization energies are \Eindreg{2} with $\eta=3.0$ a.u.\
  Polarization models are as described in the text. Damping parameters are
  specified in atomic units. Damping models either use a single damping parameter
  for all site pairs or, for the $\beta(3)$ model, use parameters that vary with
  site-pair. 
  \label{fig:H2O2-min-POL2-var-models}
  }
\end{figure}

Fig.~\ref{fig:H2O2-min-POL2-var-models} shows three polarization models
for the water dimer in its hydrogen-bonded orientation. These models differ in 
their damping parameters only. The first uses a damping parameter 
derived from the expression
\begin{align}
  \beta = \sqrt{2 I_{\rm A}} + \sqrt{2 I_{\rm B}},
    \label{eq:beta-IPs}
\end{align}
which was given by Misquitta and Stone \cite{MisquittaS08a} and is derived from the 
molecular vertical ionization energies $I_{\rm A}$ and $I_{\rm B}$. In a later 
paper on dispersion models \cite{MisquittaS08b}, these authors demonstrated that
this simple expression resulted in accurate damping models for a variety of 
systems, ranging from hydrogen-bonded complexes to van der Waals systems in 
a wide range of orientations. 
For the water dimer, using a vertical ionization energy of 0.4638 a.u.\
\cite{NIST_lias-long}, using eq.~\eqref{eq:beta-IPs} we get $\beta=1.93$ a.u.\
From fig.~\ref{fig:H2O2-min-POL2-var-models} we see that, as demonstrated in 
Ref.~\cite{MisquittaS08a}, the resulting polarization model agrees well with 
the {\em total, un-regularized} energy \Eind{2}. The agreement is particularly 
good at the dimer equilibrium separation, though at shorter separations this
model tends to {\em underestimate} \Eind{2}. Notwithstanding this seemingly 
good performance, there is considerable evidence from Sebetci and Beran \cite{SebetciB10a}
and also from our own numerical experiments that polarization models 
derived using eq.~\eqref{eq:beta-IPs} significantly overestimate the 
{\em many-body} polarization energy. By fitting to the many-body energies of
clusters of water molecules, Sebetci and Beran obtain an optimized parameter
of $\beta_{\rm opt}=1.45$ a.u., i.e., the damping needs to be considerably enhanced.

The reason for this is that the polarization
model should reproduce the true polarization energy and not \Eind{2}. 
If the model is derived to match \Eind{2}, it will implicitly include,
via the damping, some amount of the two-body charge-transfer energy.
While this is, in itself, not a significant problem for the two-body energy,
it can lead to large errors in the many-body polarization energy which will 
then include spurious two-body charge-transfer effects.
This would lead to the polarization over-binding of clusters of molecules
seen by Sebetci and Beran and discussed in Secs.\ \ref{sec:introduction} and 
\ref{sec:what-is-CT} of this article.

\begin{figure}
  \includegraphics[width=0.5\textwidth]{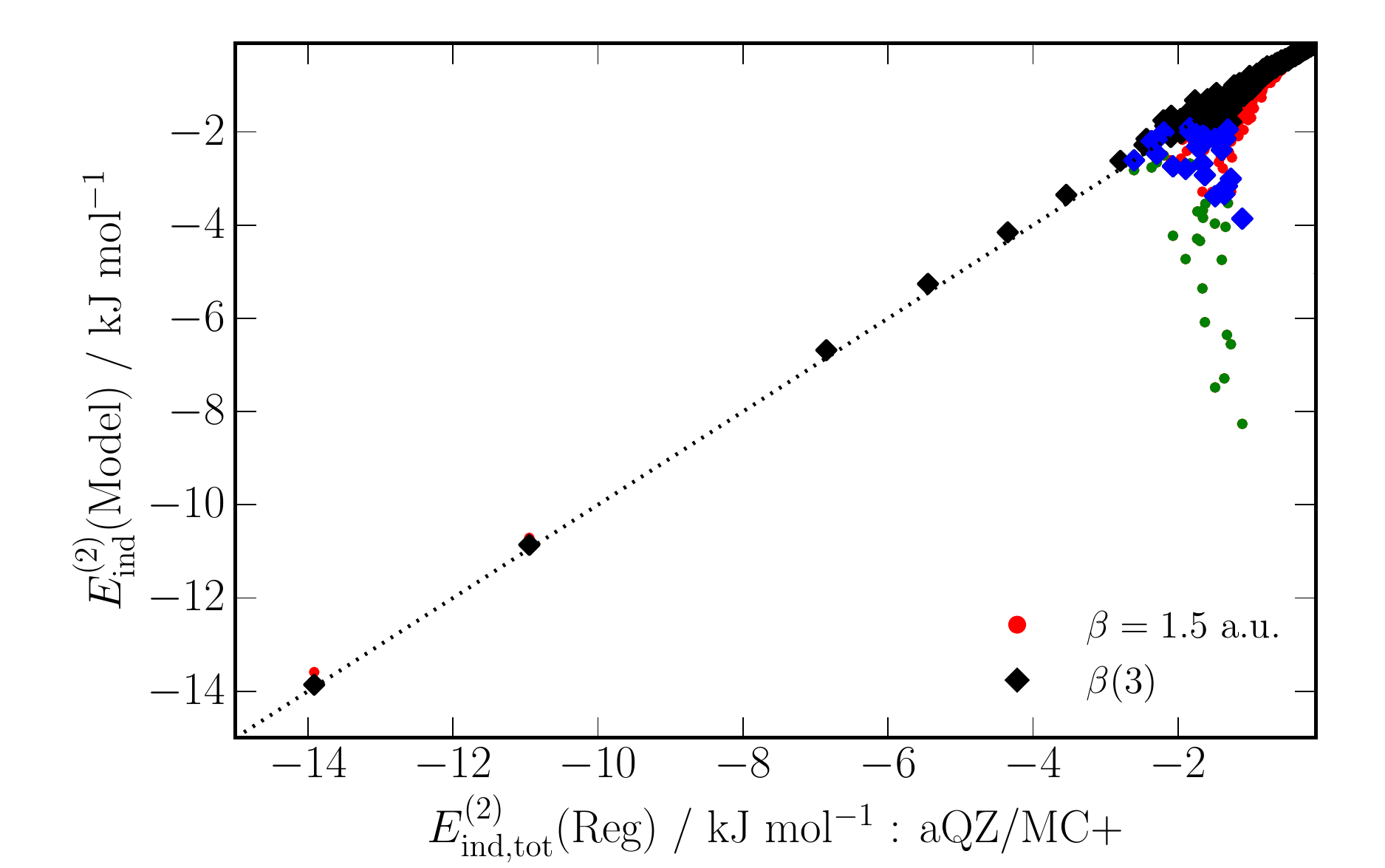}
  \caption[Water dimer in random dimer orientations: polarization models]{
  Second-order polarization energies for the water dimer in the 414 orientations
  described in the caption to fig.~\ref{fig:H2O2-E2indReg-vareta-scatter}.
  Model second-order polarization energies are plotted against 
  reference true polarization energies defined to be \Eindreg{2} with $\eta=3.0$ a.u.\ 
  calculated using the aug-cc-pVQZ/MC+ basis set.
  Polarization models are as described in the text. Damping parameters are
  specified in atomic units. 
  Two damping models are used: the $\beta=1.5$ a.u.\ (red and green dots) 
  model uses the same parameter for all site pairs and
  the $\beta(3)$ model (black and blue diamonds) uses parameters that vary
  with site-pair.
  Each set of points is divided into those with 
  $\Eexch \gt 40$ \kJmol (blue diamonds and green squares) and those 
  with $\Eexch \le 40$ \kJmol (black diamonds and red squares). The former set
  include mainly those configurations with close O$\cdots$O contacts.
  \label{fig:H2O2-POL2-var-models-scatter}
  }
\end{figure}

As discussed above, the solution to this problem is to determine the 
damping of the polarization model by fitting it to reproduce the true 
polarization energy only. If we do this using a single damping parameter,
that is, the damping parameter is independent of type of sites,
we get $\beta=1.50$ a.u.\ From fig.~\ref{fig:H2O2-min-POL2-var-models} we see
that this model results in a very good match with \Eindreg{2}, at least for
the dimer in its hydrogen-bonded orientation. However, as can be seen in 
fig.~\ref{fig:H2O2-POL2-var-models-scatter}, the agreement is not as good
for other orientations; in particular, at those with close O$\cdots$O 
separations the polarization model tends to overestimate \Eindreg{2}.
A detailed examination of the water dimer polarization energies at
orientations with close H$\cdots$H and O$\cdots$O contacts suggests that 
the damping parameter is strongly dependent on the types of the sites in
the interacting pair. A much better fit to \Eindreg{2} is
obtained with a three parameter damping model, $\beta(3)$, in which 
$\beta_{\rm OH} = 1.61$, $\beta_{\rm HH} = 1.80$ and $\beta_{\rm OO} = 1.09$
a.u.\ Of these, $\beta_{\rm HH}$ is not very well defined and can be fixed to a
relatively wide range of values. From figs.~\ref{fig:H2O2-min-POL2-var-models}
and \ref{fig:H2O2-POL2-var-models-scatter} we see that this model is significantly 
better than the simpler, one-parameter model with $\beta=1.5$ a.u.\

Even the three-parameter model exhibits somewhat larger errors for the 
dimers with close O$\cdots$O contacts (highlighted in 
fig.~\ref{fig:H2O2-POL2-var-models-scatter}).
It is possible that a proper non-linear optimization of the damping model
will result in a model that removes these discrepancies, but it is also
possible that at least some of the residual error is from the lack of
angular dependence in the damping model. These issues are currently 
under investigating.

Although it may appear that the single-parameter model with 
$\beta=1.50$ a.u.\ matches the Sebetci and Beran value of 
$\beta_{\rm opt}=1.45$ a.u., matters are not as straightforward. First of all, 
Sebetci and Beran used a somewhat different multipole and polarizability
models: their distributed multipole model had terms on the hydrogen atoms 
limited to rank 1, and in their WSM polarizability model the maximum rank was 2.
Both of these, particularly the former, result in smaller polarization energies.
Consequently, the damping need not be as large.
Indeed, using a multipole model similar to theirs and by fitting to
hydrogen-bonded dimer orientations only, we obtain a single parameter
damping parameter of 1.6 a.u.\ (a larger $\beta$ means less damping).
But these orientations are not fully representative of those found in
water clusters such as those used by Sebetci and Beran. Here the contacts
are more diverse, particularly in the more compact water clusters. 
We conjecture that the Sebetci and Beran value of 
$\beta_{\rm opt}=1.45$ a.u.\ is a compromise that is the average of the 
site-pair-dependent parameters in model $\beta(3)$ described above.
While this needs to be confirmed, we emphasise that all of these results
are consistent: the polarization damping should be much smaller than what we
would expect from eq.~\eqref{eq:beta-IPs}.

\begin{figure}
  \includegraphics[width=0.5\textwidth]{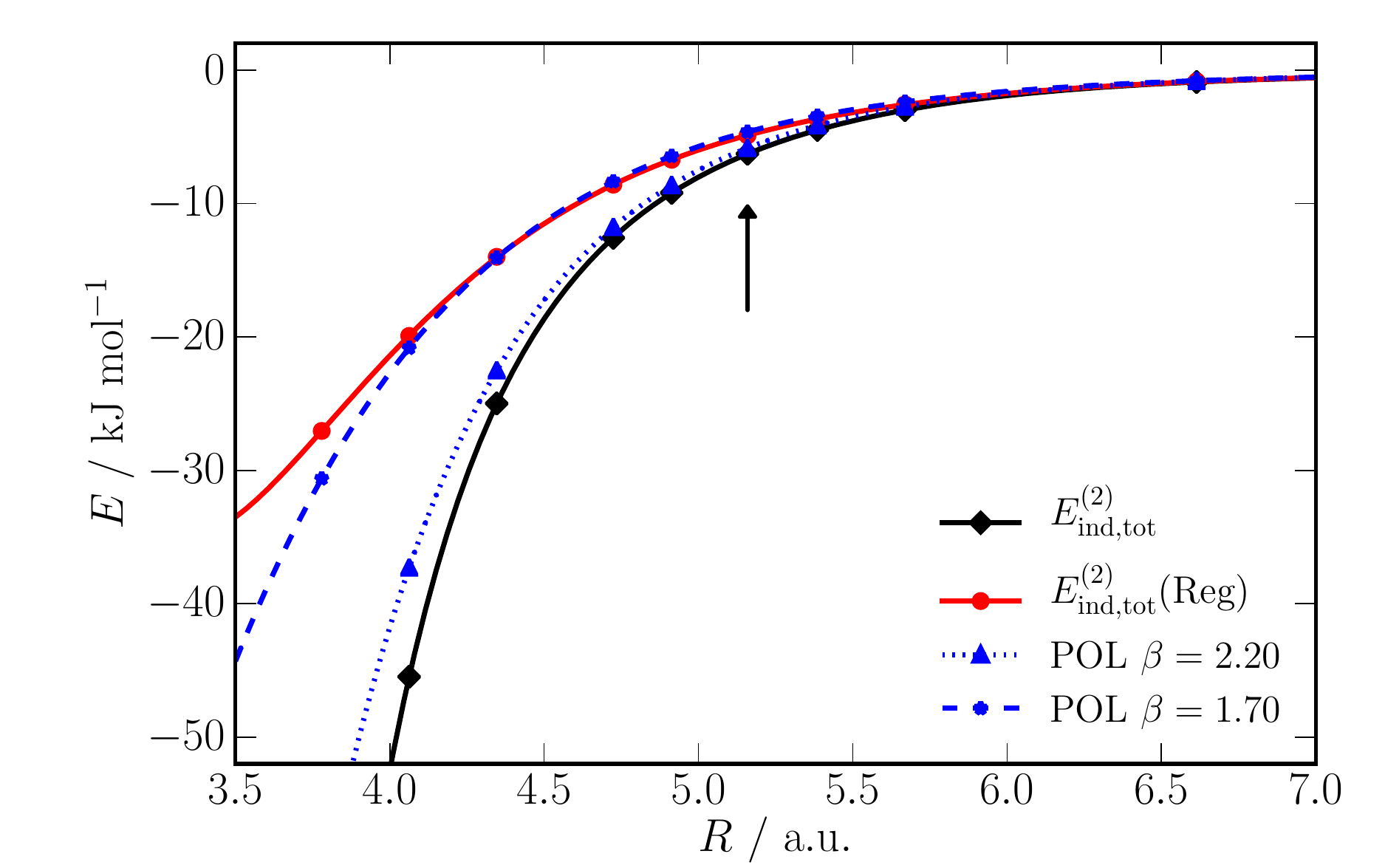}
  \caption[HF dimer in its global energy minimum configuration: 
  polarization models]{
  Second-order polarization energies for the HF dimer in its hydrogen-bonded
  orientation. For a description see the caption to fig.~\ref{fig:H2O2-min-POL2-var-models}.
  \label{fig:HF2-min-POL2-var-models}
  }
\end{figure}

\begin{figure}
  \includegraphics[width=0.5\textwidth]{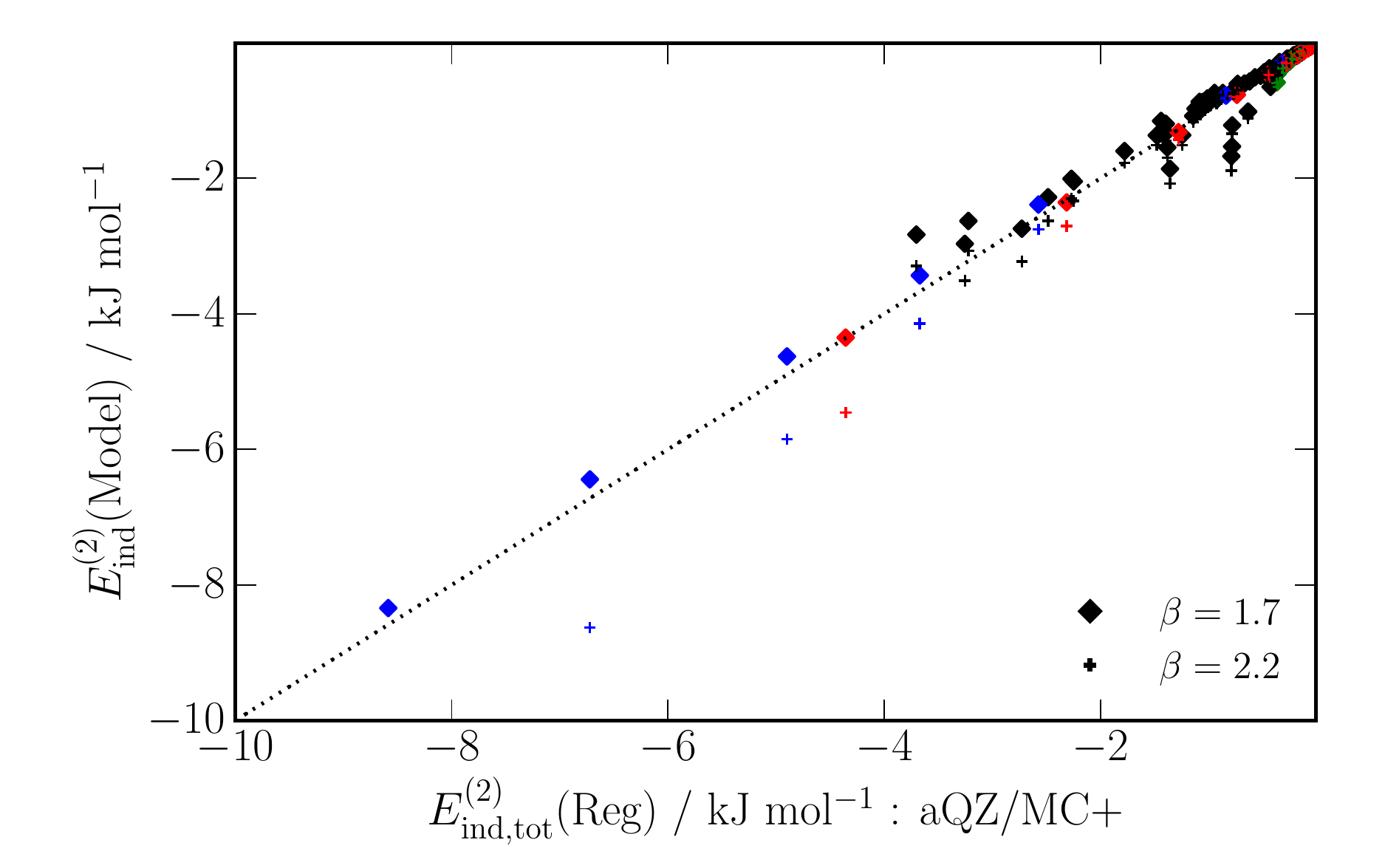}
  \caption[HF dimer in random dimer orientations: polarization models]{
  Second-order polarization energies for the HF dimer in the 159 orientations
  described in the caption to fig.~\ref{fig:HF2-E2indreg-vareta-scatter}.
  Results from two damping models are shown: $\beta=2.2$ a.u.\ (plus signs) and
  $\beta=1.7$ a.u.\ (diamonds). These data are colour-coded as follows:
  dimers with F$\cdots$H contacts (blue), H$\cdots$H contacts (red), and
  F$\cdots$F contacts (green).
  \label{fig:HF2-POL2-var-models-scatter}
  }
\end{figure}

The HF dimer exhibits many of these results. The polarization model based on
eq.~\eqref{eq:beta-IPs} and  a vertical ionization potential of
0.5669 a.u.\cite{NIST_lias-long} results in a damping parameter of 2.2 a.u.\ 
From fig.~\ref{fig:HF2-min-POL2-var-models} we see that this model
closely matches \Eind{2}; in this case the match is even better than for the
water dimer. Our second model is obtained by fitting to the true polarization
energy \Eindreg{2}. This yields a single parameter model with $\beta=1.70$ a.u.\
which, as we see from fig.~\ref{fig:HF2-POL2-var-models-scatter}, is able to 
match the \Eindreg{2} energies for other dimer orientations, albeit with a
larger scatter than we had for the water dimer. At least some of this scatter is
due to the unusual short-range mismatch in \Eindreg{2} and the polarization model
seen in fig.~\ref{fig:HF2-min-POL2-var-models}. But there is also evidence that,
as with the water dimer, an accurate polarization model would require
three different damping parameters, with the F$\cdots$F interaction needing
a considerably stronger damping. We have yet to explore this possibility fully.

Curiously, Sebetci and Beran \cite{SebetciB10a} found the original damping model
with $\beta=2.2$ a.u.\ to be suitable for cyclic clusters of HF molecules.
Exploration of this issue shows that with a multipole model similar to the
one they used (with terms limited to rank 1 on the hydrogen atoms) we obtain 
an optimized damping parameter of 2.0 a.u.\ by fitting to $\Eindreg{2}$. This
parameter should increase still more when the maximum rank of the polarization
model is reduced to 2 from the maximum of 3 which we have used. Therefore,
in this case too, we believe that our results are fully consistent with those
from Sebetci and Beran and indicate that we are indeed now able to derive 
many-body polarization models from the dimer alone.

\section{Discussion}
\label{sec:discussion}

We have presented a definition of the charge-transfer energy that is based
on interpreting the transfer of charge between molecules through the 
process of tunneling. In this view, as illustrated in fig.\ \ref{fig:CT-tunneling},
intermolecular charge transfer occurs by electron density tunneling into the
screened nuclear potential of the partner monomer. This viewpoint leads
to a simple way of defining the charge-transfer energy by 
{\em regularizing} the screened nuclear potential so as to suppress
tunneling, while still allowing classical electromagnetic polarization.

The Gaussian-type regularization we have used involves a parameter $\eta$
that has the dimensions $L^{-2}$, or, equivalently, the regularization
introduces a nuclear screening length-scale $\eta^{-1/2}$.
This is the single most important parameter in this procedure. 
{\em A priori} the screening length-scale is expected to depend on the atom type,
but this dependence has been shown to be weak.
Using a number of systems and two different procedures, we have demonstrated
that the regularization parameter, $\eta$ is very close to 3.0 a.u., that is,
the regularization occurs on a length-scale of 0.58 Bohr. 
This value of the parameter is suitable for systems involving lighter 
elements, but further work is needed to understand why this value is appropriate
and how it may be expected to change for systems containing heavier elements.

Once the value of the regularization parameter has
been fixed, the second-order charge-transfer energy, \CTreg{2}, has a 
well-defined complete-basis-set limit for all intermolecular separations. 
This strongly contrasts with the definition put forward by Stone and Misquitta
\cite{StoneM09a} which exhibits a strong basis dependence, particularly at short
separations.

With the proposed definition of the charge-transfer through regularization 
the charge-transfer energy for the water dimer and HF dimer systems exhibits
a clear double exponential character. This has been shown to stem from the strong
asymmetry in the electron donor to acceptor and electron acceptor to donor
components of the \CTreg{2} energies: the donor to acceptor process is not
only dominant, but decays more slowly with separation; indicative of a
tunneling process through a smaller barrier.
Additionally, using a partial regularization we have been able to decompose the
donor to acceptor charge-transfer energy into a primary process from the electron-rich
donor O or F atom into the electron deficient acceptor H atom, and a much smaller
secondary process from the donor O or F into the acceptor O or F.
The secondary process is an order of magnitude smaller than the primary donor to 
acceptor process.

This new definition of the second-order charge-transfer energy has been used on 
a variety of systems, including some with very strong hydrogen bonds
(H$_3$B$\cdots$CO and H$_3$B$\cdots$NH$_3$) and one doubly hydrogen-bonded
system (the pyridine dimer in its $D_{\rm 2h}$ planar configuration). 
In all cases the computed charge-transfer energy makes physical sense. 
The differences between \CTreg{2} and other definitions such as the 
Stone and Misquitta and ALMO method from Khaliullin \etal \cite{KhaliullinBH-G08}
are particularly dramatic for the most strongly bound H$_3$B complexes
for which only \CTreg{2} results in physically meaningful 
energies. In the doubly hydrogen-bonded pyridine example,
while there is a stabilisation due to the tunneling, due to symmetry
there is no nett charge transferred between the two molecules:
they remain neutral. So, in a sense, the term `charge-transfer' is
a misnomer.

Finally, we have used regularization to suppress all charge-transfer
contributions from the second-order induction energy thereby defining
a pure second-order polarization energy against which we have 
fitted the damping in the WSM polarization models. Comparisons against data
from Sebetci and Beran \cite{SebetciB10a} indicates that the resulting 
models are able to describe the many-body polarization energies accurately.
This is a major step forward in our ability to model the major part of 
the many-body non-additive energy in polarizable systems from calculations
on the dimer alone.

Based on the arguments put forward in the Introduction and
the many examples provided herein we now propose a new definition of the
charge-transfer energy:
\vspace{0.15cm}
\\
{\em 
  The process of charge transfer can be thought of as a charge
  delocalisation through electron tunneling into the screened nuclear
  potentials of the partner monomers.
  The resulting energy of stabilization is the 
  intermolecular charge-transfer or delocalisation energy.
}
\vspace{0.15cm}
\\
We suggest that it may be conceptually advantageous to term this
the {\em delocalisation} energy and reserve the term `charge-transfer' 
for the phenomenon of charge-transfer excitations studied by 
Mulliken \cite{Mulliken52}.

Though the interpretation of the CT as a tunneling process may appear
different from the usual picture of the CT as an excitation from a
donor orbital to an acceptor located on the partner monomer, we believe
that these two viewpoints are consistent.
The second-order Rayleigh-Schr\"{o}dinger perturbation
expression for the second-order induction energy given in eq.~\eqref{eq:E2indpol}
contains terms that arise from excited states that have a significant 
component at the sites of the partner monomer. For these states
to make a significant contribution to the induction energy,
the matrix elements in the numerator need to be significant (and
the energy differences in the denominator small). One way this can happen
is when the excited state has significant contributions in the regions
of the nuclei of the partners where the screened potential $V^{\rm B}$ is
significant. This is just another way of describing tunneling states.

Charge-transfer as incipient chemical bonding: if we accept the physical
picture of charge transfer as a tunnelling of electrons into the 
screened nuclear potentials of the partner monomer, then we must also
accept the view the well-established view that charge-transfer is a form of
incipient covalent bonding. 

There are several advantages to the proposed definition of the charge-transfer
energy:
\begin{itemize}
\item It leads to physically appealing definition of CT.
\item The resulting (electron) acceptor to donor and (electron) donor
to acceptor contributions make physical sense. 
\item It leads to polarization models that are applicable to many-body systems
although they are based on dimer calculations alone.
\item The method is implemented as part of regular SAPT(DFT) calculation.
\item The results are independent of basis set and true basis-set convergence
can be achieved. 
\item The method is independent of the type of basis set: it could be used,
for example, with plane-wave basis sets. Therefore it should be possible to 
use this definition to calculate the charge-transfer, or delocalisation energy,
in a variety of electronic structure programs.
\end{itemize}

Amongst the issues that need resolving is our incomplete understanding
of the origin of the regularization length-scale and how it depends 
on atom type, and the manner in which the present procedure needs to be 
extended to calculate charge-transfer effects beyond second order
in perturbation theory. Both of these issues are under current
investigation.

\section{Acknowledgements}
The author would like to acknowledge Prof Anthony J Stone and Prof Kenneth Jordan
for many stimulating and fruitful discussions, and in particular, Prof Stone for
many useful comments and suggestions and a stimulating collaboration. 
Additionally, the author thanks the School of Physics
at Queen Mary, University of London for support and the Thomas Young Centre
for providing a stimulating research platform in the London area.

\setlength\bibsep{2pt}

\end{document}